\documentclass[useAMS,usenatbib]{mn2e}

\usepackage{graphicx}
\usepackage{txfonts}
\usepackage[british]{babel}
\usepackage{latexsym}

\title[On the evolution of irradiated turbulent clouds: A comparative study between modes of triggered star-formation]{On the evolution of irradiated turbulent clouds: A comparative study between modes of triggered star-formation}
\author[Anathpindika. S. and Bhatt. H. C.]{Anathpindika. S\thanks{E-mail:
sumedh$\_$a@iiap.res.in} and Bhatt. H. C. \\
Indian Institute of Astrophysics, Bangalore 560034, India}

\begin{document}

\date{Accepted 1988 December 15. Received 1988 December 14; in original form 1988 October 11}

\pagerange{\pageref{firstpage}--\pageref{lastpage}} \pubyear{2002}

\maketitle

\label{firstpage}

\begin{abstract}
Gas within molecular clouds ({\small MC}s) is turbulent and unevenly distributed.  Interstellar shocks such as those driven by strong fluxes of ionising radiation ({\small IR}) profoundly affect {\small MC}s.  While small dense {\small MC}s exposed to a strong flux of {\small IR} have been shown to implode due to radiation-driven shocks, a phenomenon called \emph{radiation driven implosion}, larger {\small MC}s, however, are likely to survive this flux which in fact, may produce new star-forming sites within these clouds. Here we examine this hypothesis using the Smoothed Particle Hydrodynamics ({\small SPH}) algorithm coupled with a ray-tracing scheme that calculates the position of the ionisation-front at each timestep. We present results from simulations performed for three choices of {\small IR}-flux spanning the range of fluxes emitted by a typical {\small B}-type star to a cluster of {\small OB}-type stars. The extent of photo-ablation, of course, depends on the strength of the incident flux and a strong flux of {\small IR} severely ablates a {\small MC}. Consequently, the first star-formation sites appear in the dense shocked layer along the edges of the irradiated cloud. Radiation-induced turbulence readily generates dense filamentary structure within the photo-ablated cloud although several new star-forming sites also appear in some of the densest regions at the junctions of these filaments. Prevalent physical conditions within a {\small MC} play a crucial role in determining the mode, i.e., filamentary as compared to isolated pockets, of star-formation, the timescale on which stars form and the distribution of stellar masses. The probability density functions ({\small PDF}s) derived for irradiated clouds in this study are intriguing due to their resemblance with those presented in a recent census of irradiated {\small MC}s. Furthermore, irrespective of the nature of turbulence, the protostellar mass-functions({\small MF}s) derived in this study follow a power-law distribution. When turbulence within the cloud is driven by a relatively strong flux of {\small IR} such as that emitted by a massive {\small O}-type star or a cluster of such stars, the {\small MF} approaches the canonical form due to Salpeter, and even turns-over for protostellar masses smaller than $\sim$0.2 M$_{\odot}$.
\end{abstract}


\section{Introduction}
Stars condense out of cold dense clumps of molecular gas called prestellar cores which appear to be located in much larger clouds, often filamentary in shape. While the formation of stars themselves is relatively well understood, a number of questions such as the distribution of stellar masses or even the formation of prestellar cores remain amongst the unsolved questions in contemporary astronomy. It was commonly believed that stars belonged to larger clusters populated with several tens of other members (e.g. Palla 2005), but the belief is now contested by more recent observations of dozens of clusters (e.g. Guttermuth \emph{et al.} 2009). 

A better understanding of the processes leading to the assembly of potential star-forming gas probably holds the key towards unravelling issues fundamental to the theory of star-formation. Primarily, they are related to the mode of star-formation, in other words clustered as against isolated or a mixture of the two. The origin of the stellar {\small IMF}, its universal nature and the possible relation with the distribution of core masses is another issue that has baffled researchers. Also a related question is that about the time-scale on which a {\small MC} is likely to form stars. For a more detailed discussion of these points the interested reader is referred to Hartmann \emph{et al.} (2011) and Bonnell \emph{et al.} (2011).

 Furthermore, observations in a number of wave-bands, and in particular, the molecular lines at sub mm-wavelengths have now demonstrated the turbulent interiors and the non-uniform distribution of gas within molecular clouds ({\small MC}s). In fact, a significant fraction of molecular gas appears to reside in dense filamentary regions (e.g., Elmegreen 1997, Nutter \emph{et al.} 2008, Andr{\'e} \emph{et al.} 2010). However, numerical simulations over the last decade have demonstrated with considerable success the tendency of turbulent gas to generate filamentary structure on a relatively short timescale, usually smaller than a sound-crossing time (e.g., Klessen \emph{et al.} 2000, Jappsen \emph{et al.} 2005 amongst a number of other works). Though turbulence provides support against self-gravity on a global scale, locally, on scales comparable to the length of driven modes, it assembles gas that can eventually become self-gravitating which probably explains star-formation within larger filamentary structures. Young protostars compete with each-other to acquire mass from their natal pool, a scenario better-known as competitive accretion (Bonnell \emph{et al.} 2001). Evidently the formation of these dense filaments and their possible fragmentation depends on the prevalent physical conditions within the parent {\small MC}. 

Numerical simulations by Ballesteros-Paredes \emph{et al.} (1999), Heitsch \emph{et al.} (2008), and Anathpindika (2009 a,b) amongst those by several other authors, have shown that dense filaments of gas form rapidly via fragmentation of larger pressure-confined gas bodies that may form due to collisions between turbulent flows. The fragmentation itself  is a result of a process called gravoturbulent fragmentation, an interplay between the gravitational instability and hydrodynamic instabilities. These simulations have shown that the star-formation process, beginning with the assembly of gas in dense pockets, once triggered, proceeds rapidly due to the non-linear growth of instabilities. However, the fact that a MC probably spends considerable time in a relatively quiescent state before star-formation is triggered, according to some authors supports the hypothesis of slow star-formation (e.g., Krumholz \& Tan 2007).

{\small MC}s in astrophysical environments though, are also exposed to ionising radiation ({\small IR}), which is likely to profoundly affect their evolution and possibly trigger new episodes of star-formation. Large nebulae such as the Rossette nebula (e.g. White \emph{et al.} 1997; Poulton \emph{et al.} 2008), the Eagle nebula with its famous pillars of creation and trunks (e.g. Sugitani \emph{et al.} 2002), the {\small M17} nebula (e.g. Hoffmeister \emph{et al.} 2008), the star-forming cloud {\small NGC7538} (e.g. Ojha \emph{et al.} 2004), the nebula {\small SFO38} in {\small IC1396} (Choudhary \emph{et al.} 2010), and cometary globules such as {\small IC1848} (e.g. Lefloch \emph{et al.} 1997), and {\small CG12} (Haikala \& Reipurth 2010), are just a few examples of star-formation triggered by the {\small IR}. The Rosette {\small MC}, for instance, is irradiated by the nearby star-cluster {\small NGC2244} and  the radiation induced turbulence appears to have generated a network of filaments within the main cloud. More recent observations of the Rosette {\small MC} using the multi-waveband infrared cameras, PACS and SPIRE, on-board the Herschel space-observatory have revealed young star-forming regions in the junctions of these filaments (Schneider \emph{et al.} 2012).  

Secondary star-formation, i.e., cases of new episodes of star-formation triggered by the flux of {\small IR} emitted by an earlier generation of stars have also been reported in for e.g., the Orion {\small MC} (Wilson \emph{et al.} 2005). Another prognosis, observed sometimes in regions of triggered star-formation, is the increasing age of stellar population along a certain direction, also called the sequential mode of star-formation (e.g., Maaskant \emph{et al.} 2011). Hot radiation from young stars is therefore believed to play a crucial role in controlling the rate of galactic star-formation, and furthermore, it is also an important source of turbulence in galactic disks (e.g. Andrews \& Thompson 2011; Krumholz \& Matzner 2009). 

Over the last three decades several authors have studied analytically the likely fate of {\small MC}s irradiated by a strong flux of {\small IR}. Dyson (1973), for example, suggested flattening of an irradiated cloud, whereas according to Bertoldi (1989) and Bertoldi \& McKee (1990), an irradiated cloud was more likely to implode under the influence of a radiation driven shock-wave within the cloud, a phenomenon known as \emph{radiation driven implosion}. The cloud so accelerated was shown to acquire a cometary structure with a characteristic velocity of the order of a few km/s. Stars could form within this comet-shaped globule ({\small CG}) . Indeed, a number of such globules have been observationally reported as cited above. The formation of {\small CG}s due to the {\small RDI} has been demonstrated numerically for small dense clouds by for instance, Lefloch \& Lazareff (1994), and more recently by Bisbas \emph{et al.} (2011). 

 Dale \emph{et al.} (2007), on the other hand, showed that the exposure of a massive self-gravitating, turbulent {\small MC} to a flux of strong {\small IR} significantly reduced the gas-to-star conversion efficiency. While these authors considered only one choice of the strength of {\small IR}, in the present study we will consider three choices ranging from a weak to a fairly strong flux, similar to that emitted by a typical young star-cluster. The different strengths of flux are acquired by varying the temperature of the source of {\small IR} and the rate of photon-emission, as recorded in Table 1. We will then compare the evolution of an irradiated, turbulent cloud with that of a similar cloud allowed to evolve only under self-gravity (and no {\small IR}). Of particular interest is the spatial distribution of dense pockets of gas within the {\small MC} in each realisation. The plan of the paper is as follows. We shall begin by briefly discussing the expected fate of irradiated MCs for different choices of gas density.  The numerical scheme adopted for this work will be described briefly  in \S 3, and the simulations will be discussed in \S 4 before results are formally presented in \S 5. We will conclude in \S 6. 

\section[]{Evolution of an irradiated cloud}
 The physical situation is illustrated in Fig. 1, where the projection (on the plane of the sky) of the cloud facing the external source of radiation, S, and located a distance $r_{s}$ from S have been shown. The flux of {\small IR}, $d\Sigma$, received by a small area element on the surface of the cloud as that shown by the small shaded portion in Fig. 1, located at a distance $r$ from the source, S, is 
\begin{equation}
d\Sigma = \frac{\mathcal{N}_{LyC}}{4\pi r_{s}^{2}}\Big(\frac{R_{cld}}{r}\Big)^{2} \sim \frac{\mathcal{N}_{LyC}}{4\pi r_{s}^{2}}\Big(\frac{R_{cld}}{r_{s}}\Big)^{2} \ \ \mathrm{cm^{-2}\ s^{-1}},
\end{equation}
where higher order terms involving $(R_{cld}/r_{s})$ have been neglected as $r_{s}\gg R_{cld}$ in general.
The force, $F_{r}$, exerted by this flux of {\small IR} on the cloud is then
\begin{displaymath}
F_{r} = \pi R_{cld}^{2}\int_{\alpha=0}^{\alpha=\pi/2}\frac{c^{2}d\Sigma}{\xi^{(2)}n_{i}r_{s}}\sin(\alpha)d\alpha\ \mathrm{cm\ s^{-2}}, 
 \end{displaymath}
where $c$ is the speed of light, $n_{i}$, the average number density of ionised gas and $\xi^{(2)}$, the second recombination coefficient defined as
\begin{equation}
\xi^{(2)} = \frac{2.06\times 10^{-11}Z^{2}}{[T_{ion}/\mathrm{K}]^{1/2}}\phi_{2}(\beta)\ \mathrm{cm}^{3} \mathrm{s}^{-1}.
\end{equation}
 In the equation above, $\beta = \frac{1.58\times 10^{5}Z^{2}}{[T_{ion}/\mathrm{K}]}$, $T_{ion}$ is the equilibrium temperature of the ionised gas, and $\phi_{2}(\beta)$ is the second recombination coefficient (Spitzer 1978). ($Z$= 1, for a fully ionised gas).
 
Using Eqn. (1), the expression for $F_{r}$ above becomes
\begin{equation}
F_{r} = \frac{c^{2}\mathcal{N}_{LyC}}{4n_{i}\xi^{(2)}r_{s}}\Big(\frac{R_{cld}}{r_{s}}\Big)^{4}\ \mathrm{cm\ s^{-2}},
\end{equation}
Note that the denominator in this expression for radiation-force has units of velocity that may be interpreted as the rate at which the ionisation-front advances. As expected, the force becomes vanishingly small at large distances, $r_{S}$. Equation (3) for the force exerted by the flux of incident radiation permits us to define the momentum, $p_{r}$, delivered  by this flux to the gas within the cloud;
\begin{displaymath}
F_{r} = \frac{dp_{r}}{dt}.
\end{displaymath}
For a molecular cloud composed of the usual cosmic mixture the average mass of gas particles within it, $\bar{m}$ =4$\times 10^{-24}$ g. If the irradiated cloud loses gas at roughly constant velocity, $v$, then the rate of mass-loss, $\dot{\mathrm{M}}_{loss}(t)$, for this cloud can be obtained by applying the conservation of momentum to the mass lost, M$_{loss}(t)$, so that
\begin{equation}
\dot{\mathrm{M}}_{loss}(t)\equiv\frac{d\mathrm{M}_{loss}(t)}{dt} \sim \frac{1}{v}\frac{dp_{r}}{dt}\sim 0.25\frac{\bar{m}c^2\mathcal{N}_{LyC}}{r_{s}n_{i}\xi^{(2)}v}\Big(\frac{R_{cld}}{r_{s}}\Big)^{4}.
\end{equation}

The expression of the mass-loss rate here assumes that the density, $n$, within the cloud is uniform and that there is no secondary loss of the {\small IR}-flux due to absorption by dust within the cloud. In a real cloud though, a fraction of the incident flux would be lost towards heating the dust within the cloud. The heated dust will of course, re-emit in infrared wavebands. The assumption of a pristine cloud devoid of any dust, however, is unlikely to alter significantly the order-of-magnitude estimate provided by Eqn. (4). The timescale, $\tau$, over which an irradiated cloud is likely to survive is simply,
\begin{equation}
 \tau = \frac{M_{cld}}{\dot{\mathrm{M}}_{loss}(t)}.
\end{equation}

Using Eqn. (4) we calculate the mass lost by irradiated clouds of different average densities, $n$, over their respective free-fall times. The results of a demonstrative calculation performed for a photon-ionised gas of average density, $n_{i}\sim$ 10 cm$^{-3}$, and maintained at an equilibrium temperature, $T_{ion}$= $10^{4}$ K, have been plotted in Fig. 2. The test cloud for each choice of flux-strength was kept at a distance $r_{s}$ from the source such that $R_{cld}/r_{s}$= 0.01. Starting from the lowest characteristic upward, the average gas density within a cloud, $n$, increases by an order of magnitude. The three green vertical lines for the respective choice of $\mathcal{N}_{LyC}$ divides the characteristics into survival-regions, i.e., the range of {\small IR}-fluxes that a cloud can possibly endure. The minimum mass of a cloud that can survive without being photoevaporated by the incident radiation-flux is defined by the intersection of these vertical lines with the characteristics on the plot.

Small rarefied clouds sit leftward of the vertical line for $\mathcal{N}_{LyC}\ =10^{48}$ s$^{-1}$ and therefore, are likely to be photo-evaporated even by a relatively weak flux of {\small IR}. Denser and more massive clouds, to the right of this line, will be steadily photo-evaporated by a stronger flux. Photoablation is therefore the likely fate of progressively larger clouds exposed to a flux of ionising radiation. Some of the more massive clouds occupy the top right-hand corner of the plot to the right of the vertical line for $\mathcal{N}_{LyC}$ = 10$^{51}$ s$^{-1}$, and evidently need stronger fluxes before they can start losing their mass.

The test cloud used in the simulations below ($n\sim 10^{4}$ cm$^{-3}$,$R_{cld}$ = 1 pc; see \S 3.4 below), lies on the characteristic just below the uppermost in the region between the fluxes corresponding to 10$^{49}$ s$^{-1}$ and 10$^{51}$ s$^{-1}$, and therefore, is expected to steadily photo-ablate. In fact, the mass-loss rate for this test cloud according to Eqn.(4) above is a few times $10^{-4}$ M$_{\odot}$ yr$^{-1}$ for $\mathcal{N}_{LyC}\sim\ 5\times10^{49}$ s$^{-1}$,  and $R_{cld}/r_{s}\sim$ 0.08. The mass-loss rate for this cloud suggests, it is likely to survive at least an order of magnitude longer than its free-fall time.

\begin{figure}
  \vspace{20pt}
   \includegraphics[width=7cm, angle=0]{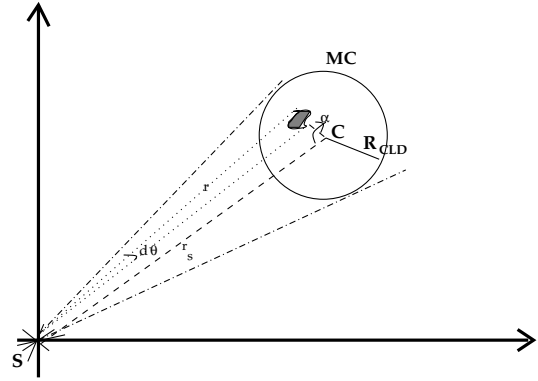}
   \caption{A  projection showing the location of the molecular cloud ({\small MC}), of radius R$_{cld}$, relative to the source of ionisation, $S$. The amount of flux received by the shaded area element on the surface of the cloud is given by Eqn. (1) in the main text. }
   \end{figure}

\begin{figure}
  \vspace{20pt}
   \includegraphics[width=7cm, angle=270]{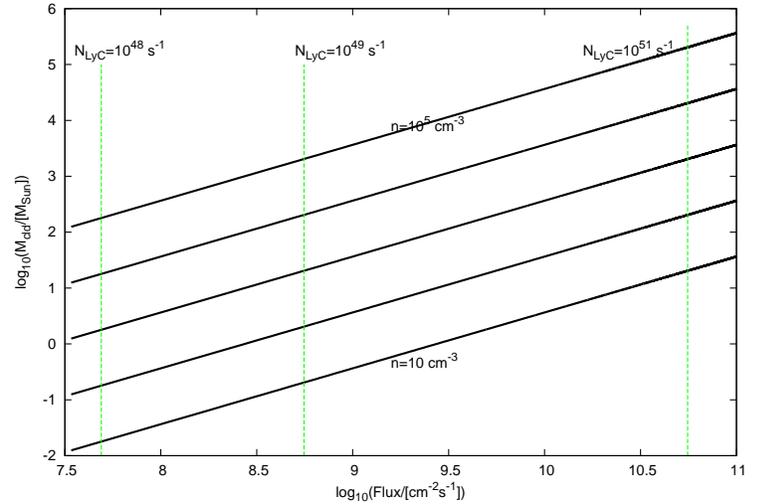}
   \caption{A simple classification scheme for {\small MC}s that enables us to predict its evolution when exposed to a flux of {\small IR}; the calculation assumes $n_{i}$ = 10 cm$^{-3}$, $T_{ion}$ = 10$^{4}$ K and $r_{s}/R_{cld}$=12. The three choices of $\mathcal{N}_{LyC}$ used in the present work are shown by green vertical lines. The intersection of these lines with the cloud mass-flux characteristics defines the minimum mass required for a cloud of given density to survive the incident flux of {\small IR}. \emph{See text for description.}} 
   \end{figure}

\section[]{Numerical Method}
\subsection{Smoothed Particle Hydrodynamics ({\small SPH})} The {\small SPH} algorithm was first introduced by Gingold \& Monaghan (1977), and Lucy (1977) to handle complex problems in astrophysical context and other areas of computational fluid dynamics (e.g. Monaghan 2005). We use our numerical hydrodynamics code, {\small SEREN}, an open-{\small MP} parallised algorithm that has been extensively tested for numerous applications of astrophysical interest (Hubber \emph{et al.} 2011). The fundamental quantity in an {\small SPH} code, the density of particles, is calculated using the \emph{gather} technique. The search for nearest neighbours of an {\small SPH} particle, and the calculation of the net force (gravity+hydrodynamic) on an {\small SPH} particle is done by distributing particles on the Barnes-Hut tree, with a cell-opening angle of $\sim$0.2; quadrupole moments of distant cells in the tree structure are also included in the calculation of forces. We have used the standard prescription for {\small SPH} artificial viscosity with the viscous parameters having values (0.1,0.2). 

The effects of ionising radiation are included through a ray-casting algorithm devised by Bisbas \emph{et al.}(2009), this scheme is an inherent  part of {\small SEREN}. The basic features of this algorithm are : it - (a) determines the position of the ionisation front ({\small IF}) at a given instant of time, and (b) assigns appropriate temperatures to the {\small SPH} particles. Of the two, (a) is achieved via the {\small HEALP}ix algorithm (G{\'o}rski \emph{et al.} 2005). Individual rays originating from the source of {\small IR} that trace the direction of propagation of the radiation are distributed uniformly on a sphere by {\small HEALP}ix. Rays are distributed at different levels, $l$, between 0 and 7 so that there are $12\times 4^{l}$ rays on any level. The angular resolution of the {\small IF} is determined by the value of $l$. The maximum intensity, $I_{max}$, at the {\small IF} from Eqn. (6) is
\begin{equation}
I_{max}\equiv \frac{m^{2}\mathcal{N}_{LyC}}{4\pi\xi^{(2)}},
\end{equation}
where $m=\frac{\bar{m}}{X}$, is the mean mass of a gas molecule and $X$=0.7, and $\xi^{(2)}$ is the second recombination coefficient defined by Eqn. (2) above.

 The integral form of the above expression for intensity at a point $r$ is,
\begin{equation}
I(r) = \int_{r} \rho^{2}(r\mathbf{\hat{e}})r^{2}dr,
\end{equation}
where $\mathbf{\hat{e}}$ is the unit vector in the direction of $r$. Equation (7) is solved iteratively to determine $r$ such that $I(r)=I_{max}$, while calculating the {\small SPH} density at each point of the iteration. 
\subsection{Thermodynamics}
All points satisfying the inequality $I(r) < I_{max}$, lie within the HII region and are therefore assigned a temperature, T$_{ion}$, the equilibrium temperature of the HII region. While the equilibrium temperature probably depends on the coupling between matter and radiation (e.g., Genel \emph{et al.} 2012), in the present study, T$_{ion}$, is assigned a fiducial value of 8000 K. Neutral gas within the irradiated cloud is treated isothermally and maintained at a temperature, T$_{n}$ = 20 K. The algorithm also includes a temperature smoothing scheme that alleviates the discontinuity arising due to a sharp difference in the temperature of the ionised and neutral gas particles. The smoothing scheme is essentially a Taylor-series approximation to the first order. 

\begin{table}
 \centering
 \begin{minipage}{80mm}
  \caption{Listed below are the choices of ionising flux and the temperature of the source used in various test cases (Spitzer 1978). The temperature of the ionised gas is maintained at, T$_{ion}$ = 8000 K in cases 2, 3 and 4. }
  \begin{tabular}{@{}rcc@{}}
  \hline
  \hline
Serial & $\frac{\mathcal{N}_{LyC}}{[\mathrm{s}^{-1}]}$; $\frac{\mathrm{T}_{star}}{[K]}$ & $\Sigma_{\mathcal{N}_{LyC}}$ \\
No.  &   &  [cm$^{-2}$s$^{-1}$]\\
\hline  
   1  & No ionising radiation & \\ 
\hline
   2   & 4$\times 10^{48}$;  &  $1.10\times 10^{7}$  \\
      &  36,000 K &  \\
\hline
   3  & 5$\times 10^{49}$;   &  $1.37\times 10^{8}$ \\
      & 47,000 K & \\
\hline
   4  & 7$\times 10^{51}$;  &  $1.93\times 10^{10}$  \\
      &  51,000 K & \\ 
\hline
\end{tabular}
\end{minipage}
 $\mathcal{N}_{LyC}, T_{star}, \Sigma_{\mathcal{N}_{LyC}}$, are  respectively the rate of ionising photons, the temperature of the source star and the flux of ionising radiation
\end{table}

\subsection{Sink particles}
An {\small SPH} particle with density higher than the threshold, $\rho_{thresh} \sim 10^{-15}$ g cm$^{-3}$, is replaced by a special type of particle, the sink, following the prescriptions made by Bate \& Burkert (1997). Apart from the density threshold, a sink particle is also has a radius which in the present simulations is set as 2.5 times the {\small SPH} smoothing length of a prospective sink particle. {\small SPH} particles bound gravitationally to the sink are accreted by it. The sink particle in the present numerical exercises represent a protostellar object. The minimum mass of an {\small SPH} particle, M$_{min} \ = \ {\mathrm N}_{neibs}\Big(\frac{\mathrm{M}_{cld}}{\mathrm{N}_{tot}}\Big)$, where $\mathrm{N}_{neibs}\ = \ 50$ and $\mathrm{N}_{gas}\ =\  550,000 $, are respectively the number of nearest neighbours of an {\small SPH} particle and the number of particles within the cloud. With this choice of {\small SPH} parameters the ratio of the thermal Jeans mass, $\mathrm{M}_{J}(\bar{\rho}\sim 10^{-16}\mathrm{g\ cm}^{-3},T_{n}=20\ \mathrm{K})$ against the minimum mass, M$_{min}$, is 6 which satisfies the Bate-Burkert criterion of resolving the gravitational instability. 

\subsection{Initial conditions}
The test cloud used in the present study was modelled as a uniform-density sphere of unit mass and radius and was assembled by randomly positioning particles within it. Since the randomness in positioning particles is inherently associated with Poissonian noise, we first evolved it for a fraction of a sound-crossing time allowing the  numerical viscosity to dissipate the spurious noise. This procedure settled the initial assembly of particles which was then rescaled to the desired dimensions. The mass and radius of the cloud are respectively, $M_{cld} = 400$ M$_{\odot}$, and radius, $\mathrm{R}_{cld}$ = 1 pc, where as the gas within it was maintained at a uniform temperature, ${\mathrm T}_{cld}$ = 20 K. A supersonic, Gaussian velocity field with random amplitudes and a relatively steep power-spectrum, $P(k)\propto k^{-4}$,  was then superposed on this cloud \footnote{$k \& P(k)$, are respectively the wave-vector and the power in the wave-vector space}. This velocity field had an initial Mach number of 10 and was first set up on a grid with 128$^{3}$ cells in the $k$-space. The test cloud was then mapped on the $k$-space and velocity components for individual particles were calculated via interpolation. Then a random phase was added to each velocity component of a particle while transforming it back into the Cartesian-space. The cloud with its turbulent velocity field was then allowed to evolve under self-gravity; this is the first case listed in Table 1. This cloud soon developed filamentary structure as shown in the rendered density plot of Fig. 3 ($t\sim 0.05$ Myr), which was then used as the initial condition in the remaining 3 cases, listed 2 through to 4 in Table 1. The cloud age was not reset to zero at the time of introducing the source of {\small IR} in these latter cases.  Each test case discussed here took a little over 5000 CPU hours and was run on the {\small HYDRA} supercomputing cluster at the Indian Institute of Astrophysics. The cluster is composed of the Intel Xeon-5675 processor. Simulations were allowed to form 150 sink particles before they were terminated.

\begin{figure}
 \vspace{20pt}
  \includegraphics[angle=270, width=8cm]{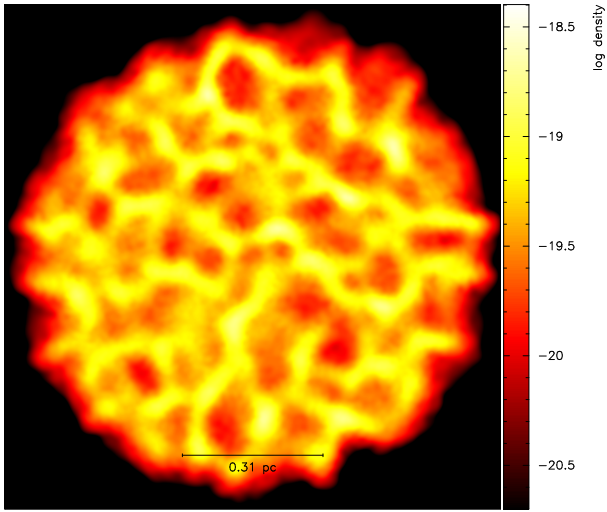}
      \caption{A rendered density plot, at $t = 0.2t_{ff}\sim$ 0.05 Myr,
 showing the mid-plane of an initially turbulent MC. Injected turbulence readily generates dense filaments within the cloud, as can be seen in this plot. This cloud is then used as the initial condition for the remaining test cases when the source of IR is turned on ($t_{ff}\sim$ 0.27 Myr).}
\end{figure}

\section[]{Numerical Simulations}
\subsection{Turbulent cloud without a source of ionising radiation}

\begin{figure}
 \vspace{20pt}
  \includegraphics[angle=270, width=9.cm]{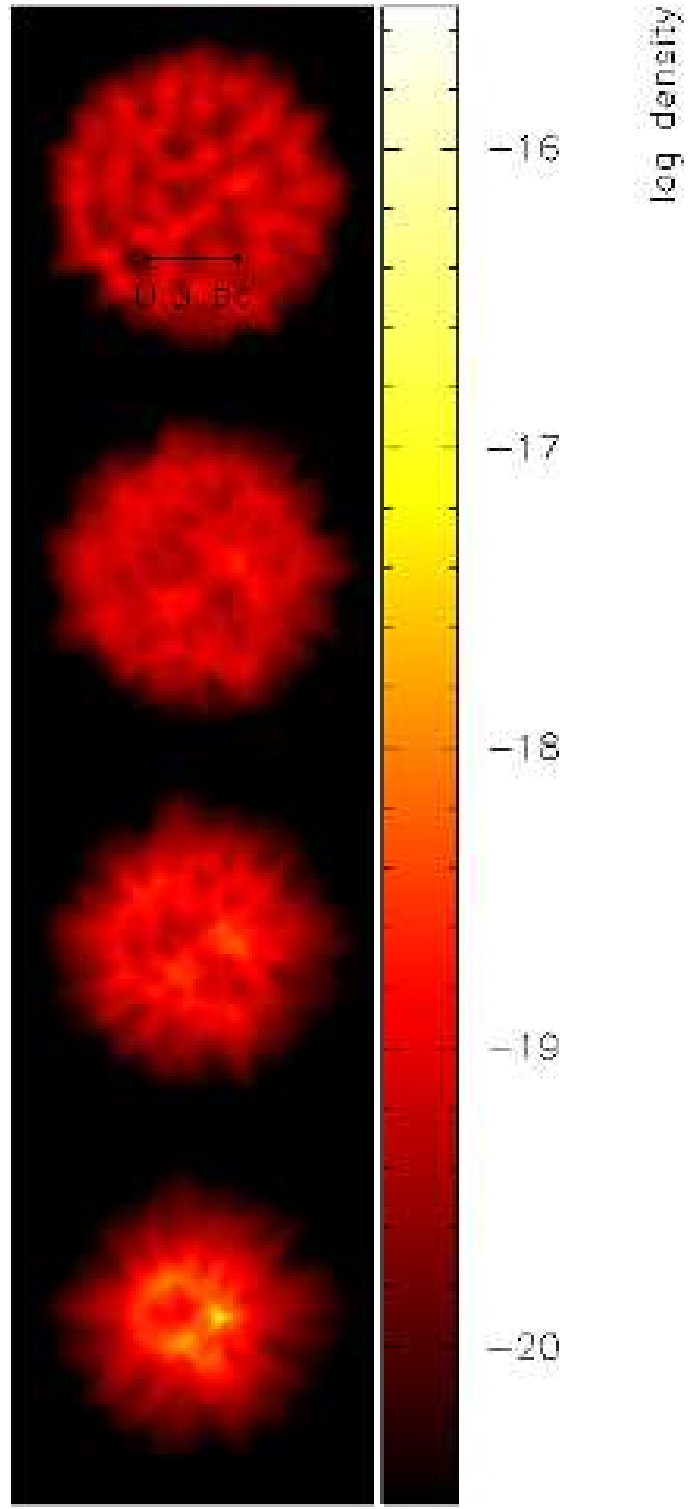}
      \caption{A sequel of column density plots showing the mid-plane of the turbulent cloud at different epochs; $t$ = 0.1,0.15, 0.2 $\&$ 0.25 Myr, for images in respective panels from the top to bottom ($t_{ff}\sim$ 0.27 Myr). Turbulence within the cloud generates filamentary structure that can be seen in the figure on each panel.}
\end{figure}

Collisions between random gas flows within a turbulent cloud dissipate turbulent energy and generate filamentary structure within the cloud as can be seen in the renders plotted in Figs. 3 and 4.  The initial velocity field, with a power-spectrum considerably steeper than the Kolmogorov type, was injected between wavenumbers 1 and 8. As a result turbulent energy cascades down via shocks from 
 larger spatial scales. Respective plots in the panels of Fig. 4 show the rendered density images of the mid-plane of the cloud at successive epochs. After a little over one free-fall time the original cloud collapses to a centrally located dense filament with a few arterial extensions. The advanced stage of the collapsed cloud can be seen more closely in the rendered plot of Fig. 5 where positions of a few protostellar objects have also been marked. As expected, the formation of protostars proceeds along the dense filaments.

\begin{figure}
 \vspace{20pt}
  \includegraphics[angle=270, width=8cm]{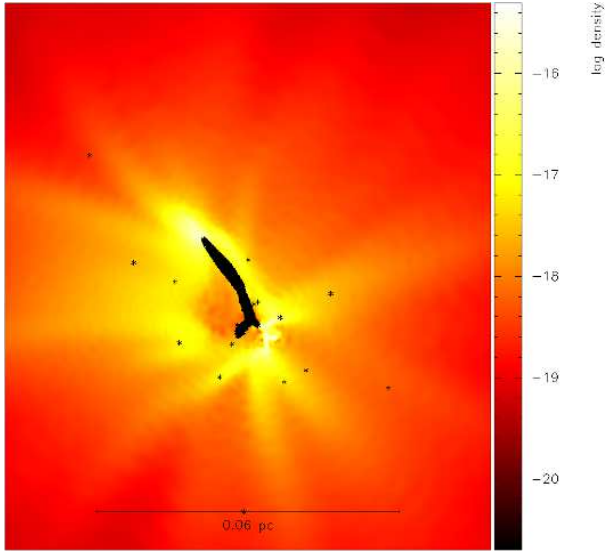}
      \caption{A column density plot showing an advanced stage in the evolution of the turbulent cloud in case 1; $t$ = 0.31 Myr$\sim$ 1.2 $t_{ff}$. The position of the sink particles within the collapsed cloud has been marked with * on this plot. Note that sinks represent protostellar objects.}
\end{figure}

\subsection{Irradiated turbulent cloud}
\begin{figure}
 \vspace{20pt}
  \includegraphics[angle=270, width=9.cm]{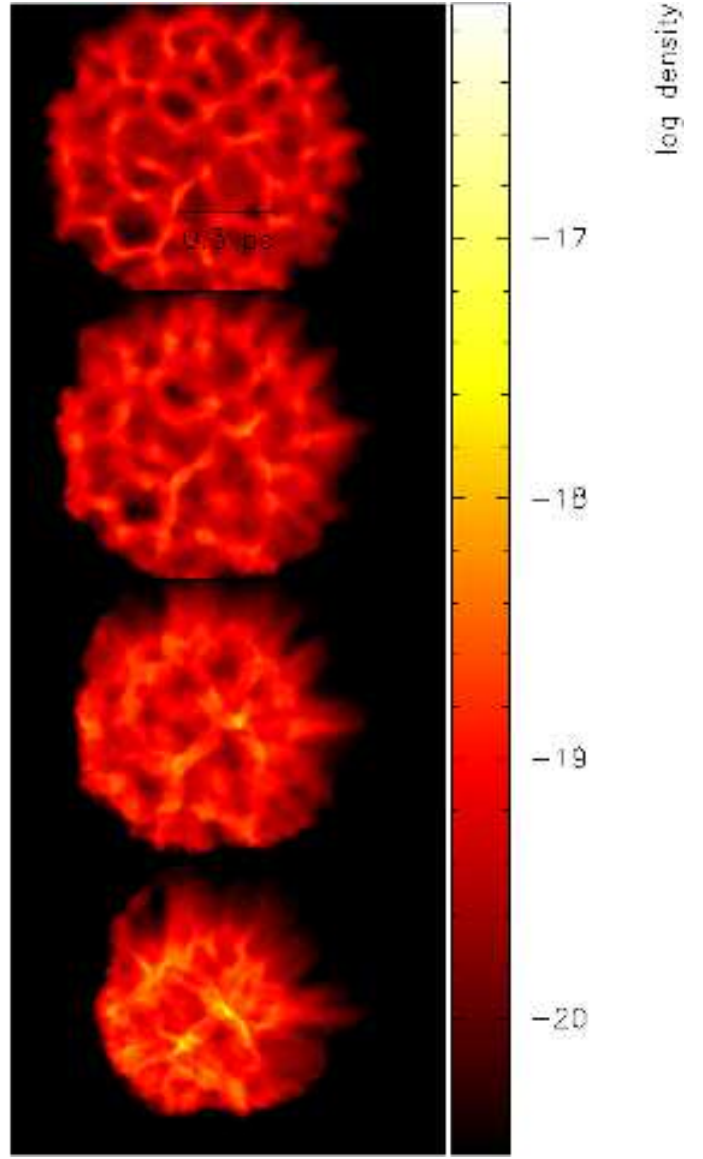}
      \caption{The mid-plane of the cloud in case 2 irradiated by a weak flux of {\small IR} is shown in this sequel of column density plots. A faint trunk that eventually evaporates due to the flux of {\small IR} can be seen in the lower left corner of the second panel. While exposure of the cloud to the incident flux of {\small IR} causes the cloud to loose mass, the radiation-induced shocks produce dense filamentary structure within the cloud; especially see the plot in the lower panel ($t$ = 0.216 Myr $\sim$0.8 $t_{ff}$).}
\end{figure}

Exposure to a flux of ionising radiation commonly causes the molecular cloud to lose mass via photo-ablation. Unlike small, dense clumps which when exposed to {\small IR} implode to form comet-shaped globules, the larger clouds such as the one considered in the present study develops sporadic pockets of dense gas, often filamentary and eventually spawn stars. In this next set of 3 simulations we endeavour to study the effect of varying the strength of {\small IR} on the evolution of a {\small MC}.

\begin{figure}
 \vspace{20pt}
  \includegraphics[angle=270, width=8cm]{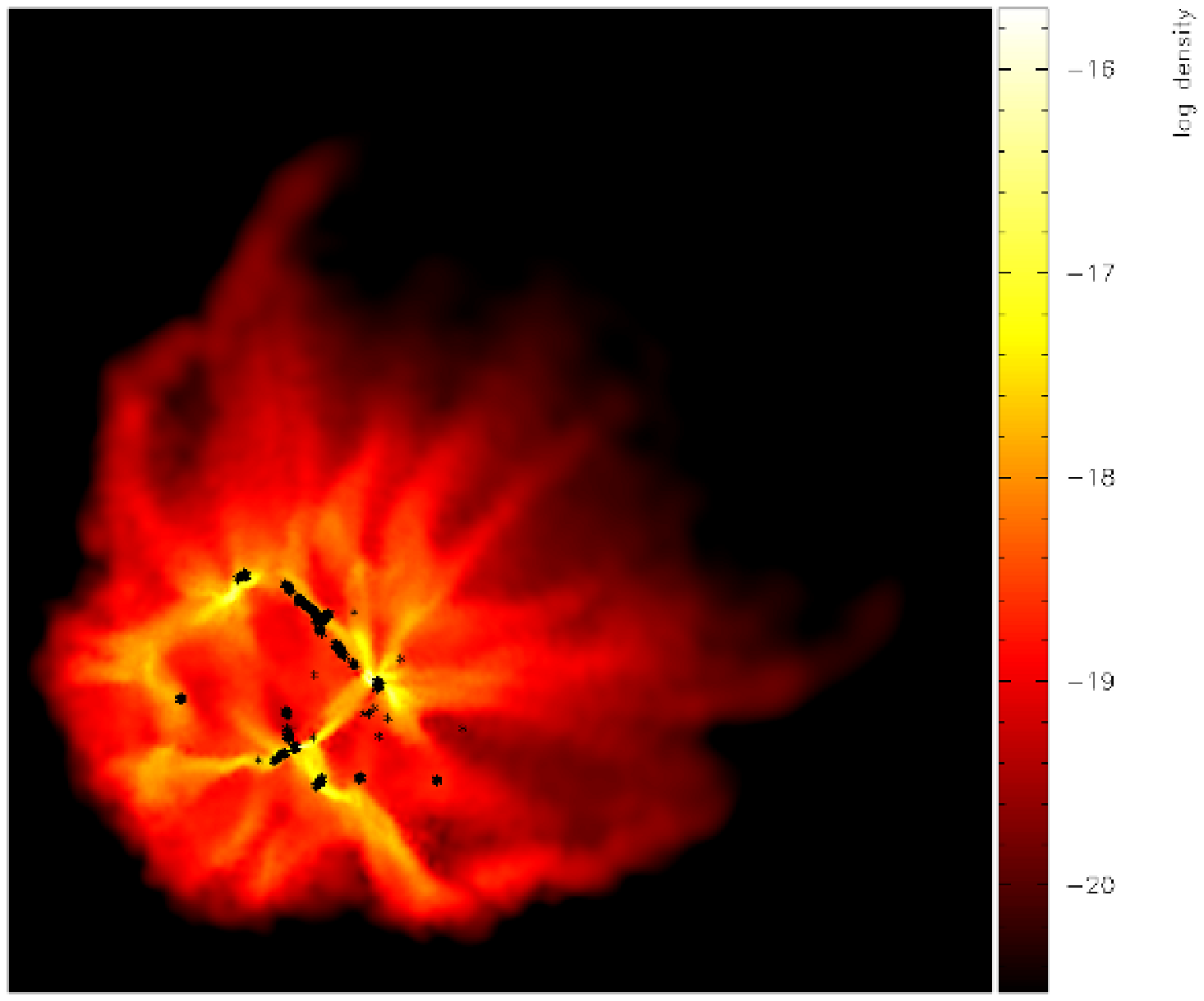}
      \caption{A column density plot showing the picture in the lowest panel of Fig. 6. Dense filaments within the irradiated cloud are readily visible in this plot and a few sink particles (i.e., protostellar objects) have been marked with *, which unsurprisingly, are located along the filaments ($t\sim$ 0.24 Myr).}
\end{figure}

\begin{figure}
 \vspace{20pt}
  \includegraphics[angle=270, width=8cm]{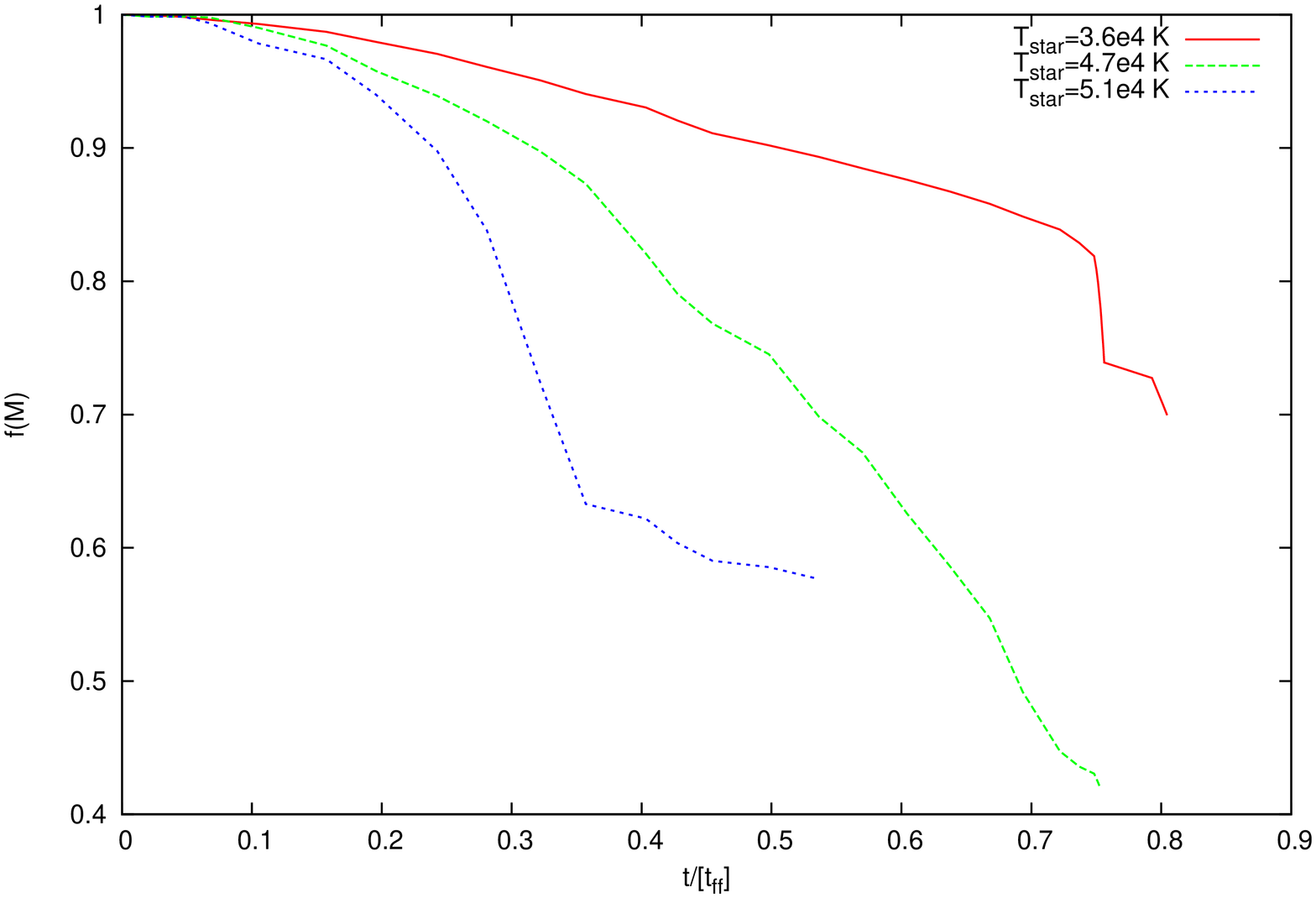}
  \includegraphics[angle=270, width=8cm]{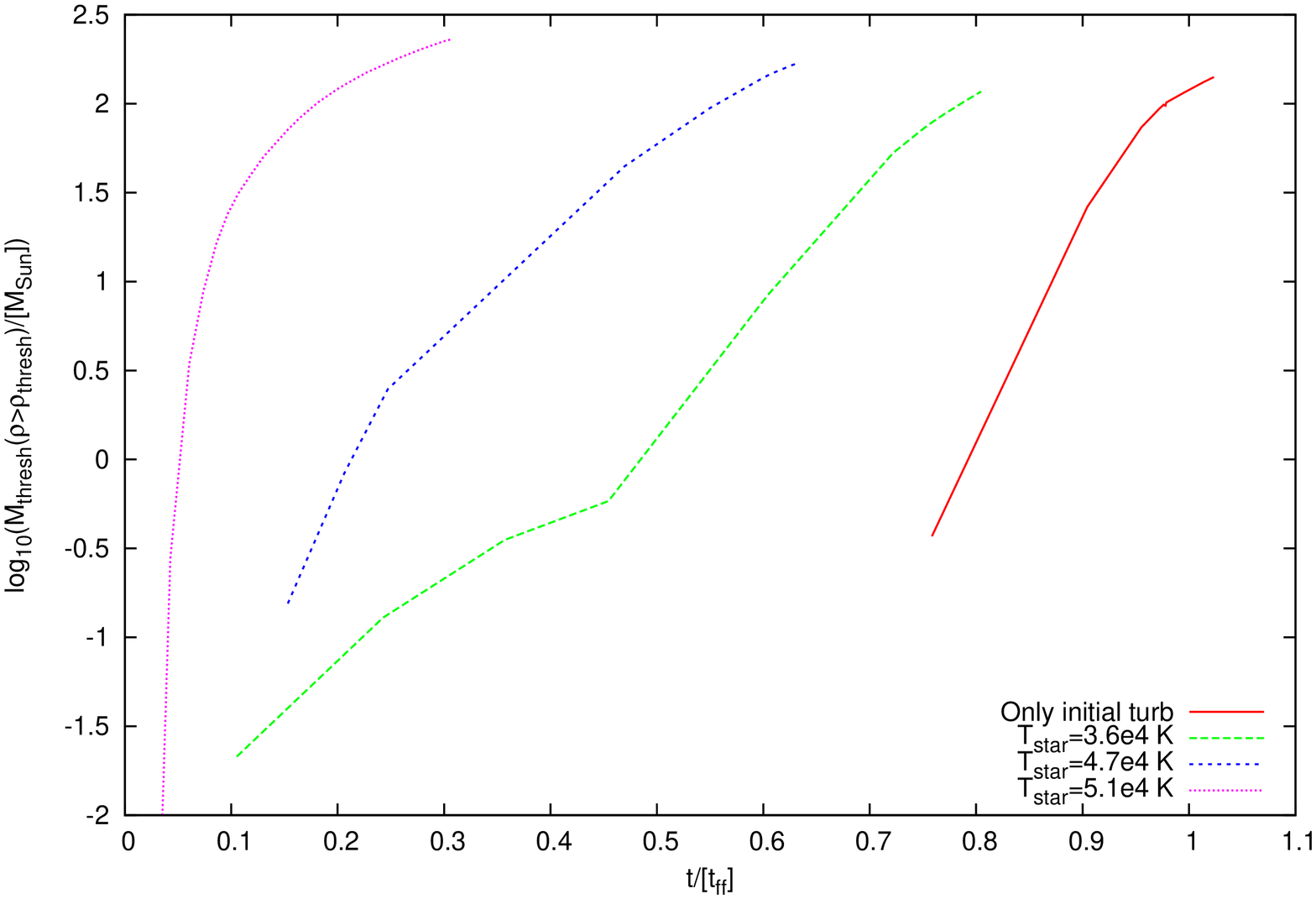}
      \caption{\emph{Top-panel} : Plot showing the fractional mass, $f$, of the irradiated cloud retained in each of the three realisations. \emph{Lower-panel} : The mass of gas within the cloud above the density threshold, $\rho_{thresh}\sim 10^{-18}$ g cm$^{-3}$, has been shown at different epochs of time. Relatively stronger shocks, such as the ones in cases 3 and 4, assemble material in the dense phase on a much shorter timescale.}
\end{figure}

\subsubsection{Weak ionising radiation}
The exposure of a {\small MC} to a relatively weak flux of {\small IR}, listed as case 2 in Table 1, produces a C-shaped ionisation front ({\small IF}), on the surface of the cloud exposed to the source of {\small IR}. The incident flux of {\small IR} also shocks the pre-existing filaments within the cloud, visible in the plots on the upper two panels of Fig. 6. Interestingly, the shocking of a clump in one of these filaments also generates a small, faint elephant trunk-like extension that can be seen in the lower left-hand corner of the second panel of this figure. The growth of this trunk is probably the result of a backward re-expansion of material from a shocked clump in a filament within the cloud. This possibility has previously been demonstrated in numerical simulations by for instance, Williams \emph{et al.}(1999), and more recently by Gritschneder \emph{et al.} (2009,2010). Protruding trunks are commonly found in star-forming clouds and some even harbour a YSO in the head region such as those reported in the famous M16 (Eagle) nebula (e.g., White \emph{et al.}1999). However, the trunk observed here ablated soon after its appearance. Turbulence injected by the flux of {\small IR} further assists the formation of filaments within the cloud, the late phase of which, with a network of filaments and star-forming sites within it can be seen in the rendered density plot in Fig. 7. Also marked on this rendered plot are the positions of protostars that appear aligned with dense gas filaments.

  The fraction, $f\equiv \mathrm{M}_{ret}(t)/\mathrm{M}_{cld}$, of the irradiated cloud retained has been plotted in the upper panel of Fig. 8. The fraction of the cloud lost is therefore $(1-f(t))$. Exposure to a weak flux of IR causes the cloud in this case to lose $\sim$30\% of its original mass and so a large portion of the cloud is still retained. The residual cloud evolves like the turbulent cloud discussed in \S 4.1 above, although the network of filaments within it, as can be seen in Fig. 7 and the lowest panel of Fig. 6, is significantly more branched in stark contrast to the filament seen in Fig. 5. The incident flux of a weak {\small IR} appears to drive many more modes. The simulation was terminated after about 0.9 free-fall times ($t$ = 0.24 Myr), by which time 150 protostars had formed within the cloud.

\subsubsection{A relatively strong ionising radiation}

With a stronger flux of {\small IR}, the {\small MC} in this case, listed 3 in Table 1, loses mass at a much higher rate than that observed in the previous realisation. Consequently, after $t=0.8 t_{ff}\sim 216$ kyrs, a little over 40 \% of the original cloud is lost via photo-ablation. Filaments in the original cloud, as is evident from the sequel of rendered density plots in Fig. 9, are shocked by the incident flux of {\small IR}. This flux also produces a thin, dense shocked-shell on the surface of the irradiated cloud. The surface of the {\small IF} as can be easily seen, is corrugated. A closer examination reveals its wiggly nature which resembles features of the well-known thin-shell instability ({\small TSI}). The {\small TSI}, as the name suggests, is triggered in thin shells confined by ram-pressure either on one or both faces. The instability grows due to the transfer of momentum between perturbed regions of the shell as demonstrated by Vishniac (1983), through a detailed perturbation analysis of the problem. The dynamically unstable nature of the irradiated surface has also been discussed by Williams (2002), who demonstrated the appearance of corrugations on the surface of the {\small IF}. However, it is unclear if these later corrugations are also the manifestations of the {\small TSI}, though for the moment we do not attempt to distinguish between the two.

We therefore use a previously derived expression for the length of the fastest growing mode, $\lambda_{fast}$, to verify if the instability is indeed resolved in this simulation. The length of the fastest growing mode in a shocked layer of uniform surface density, $\sigma_{0}$, calculated by Anathpindika (2010) is
\begin{displaymath}
\lambda_{fast} = \frac{2\pi}{k_{fast}},
\end{displaymath}
where 
\begin{equation}
k_{fast} = \frac{\pi G\sigma_{0}}{\Big[a_{0}^{2} - \frac{P_{E}}{\rho_{s}}\Big(1-\frac{R_{s}}{R_{2}}\Big)^{-1}\Big]}.
\end{equation}
This mode grows on a timescale, $t_{growth}\sim \lambda_{fast}/a$; $a=(k_{B}T_{gas}/\bar{m}_{H})^{1/2}$ is the sound speed  and, $T_{gas}$, the average temperature of the gas in this layer, which is $\sim 10^{4}$ K. The radius of the irradiated cloud, $R'_{2} = R_{cld} + dR$, where $R_{cld}$ is the radius of the original cloud and $dR$ is the thickness of the shell. In the present case,  $dR/R'_{cld}\sim 10^{-2}$, and  $P_{E}/\rho_{s}\sim a_{HII}^{2}$ is the square of the sound-speed in ionised gas. Plugging in the appropriate values we get, $\lambda_{fast}\sim $0.05 pc, and $t_{growth}\sim $0.22 Myr, which agrees with the timescale on which wiggles first appear on the surface of the shocked-shell.

The ratio of $\lambda_{fast}$ to the average {\small SPH} smoothing length, $\mathcal{X}$, is a good indicator of the spatial resolution, which in the present case is 3. It is therefore likely that the instability has been only barely resolved here. Better resolution of the instability demands at least an order of magnitude increase in $\mathcal{X}$, which transforms in to about 3 orders of magnitude increase in the number of {\small SPH} particles. This is computationally demanding. However, since we are not concerned about the detailed modelling of this instability, the present choice of spatial resolution is sufficient. It is therefore clear that radiation-induced shock produces pockets of dense gas along the irradiated surface, and further, a few other isolated pockets of dense gas can also be seen along the shocked filaments in the interior of the cloud. The rendered density plot in the left-hand panel of Fig. 10 shows the irradiated cloud more closely along with the location of sink particles on it. As with the earlier case, this simulation was also terminated at $t\sim 0.22$ Myr, by which time the cloud had formed 150 sink particles. We note that the sink-formation timescale in cases 2 and 3 is therefore mutually comparable.

\subsubsection{Strong ionising radiation}
In this final test case, listed number 4 in Table 1, the flux is the strongest with source characteristics similar to those of a typical young star-cluster. The irradiated cloud in this case evolves in a manner similar to that observed in the previous case, albeit on a much shorter timescale and has a much higher mass-loss rate, as also reflected by the plots in Fig. 8. Analogous to the previous 2 cases, exposure to a flux of {\small IR} shocks the surface and the interiors of the cloud. Some of the clumpy features near the surface briefly appear to develop trunk-like features that  are stymied soon due to rapid photo-ablation of the surface. The surface of the irradiated cloud, as in case 3, also shows evidence of the thin-shell instability discussed previously in \S 4.2.2. Since the broad features of the irradiated cloud in this case match with those in the previous realisation, we only show the final state of the cloud in the right-hand panel of Fig. 10. Unlike the irradiated cloud in the previous case the initial star-formation in this case is confined predominantly to the shocked shell

\begin{figure*}
\vbox to 70mm{\vfil
 \includegraphics[angle=270, width=16.cm]{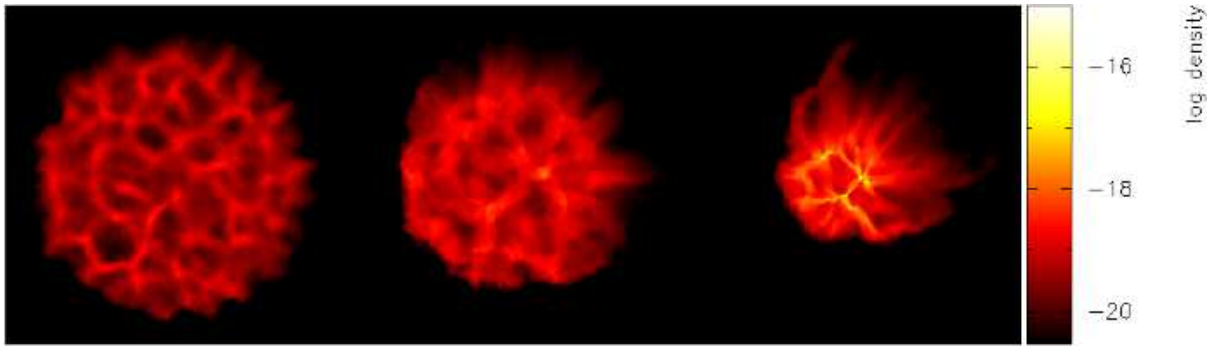}
 \caption{Mid-plane of the irradiated cloud in case 3 has been shown in these rendered column density plots. Exposure to a relatively strong flux of IR ablates the cloud while also shocking the dense filaments within it as can be seen in the plots on each panel; $t\sim 10^{5}$, 2$\times 10^{5}$ and 2.1$\times 10^{5}$ yrs ($\sim 0.7 t_{ff}$), for plots in the left-, central and the right-hand panels respectively. } \vfil} \label{landfig}
\end{figure*}

\begin{figure*}
\vbox to 90mm{\vfil
 \includegraphics[angle=270, width=8.cm]{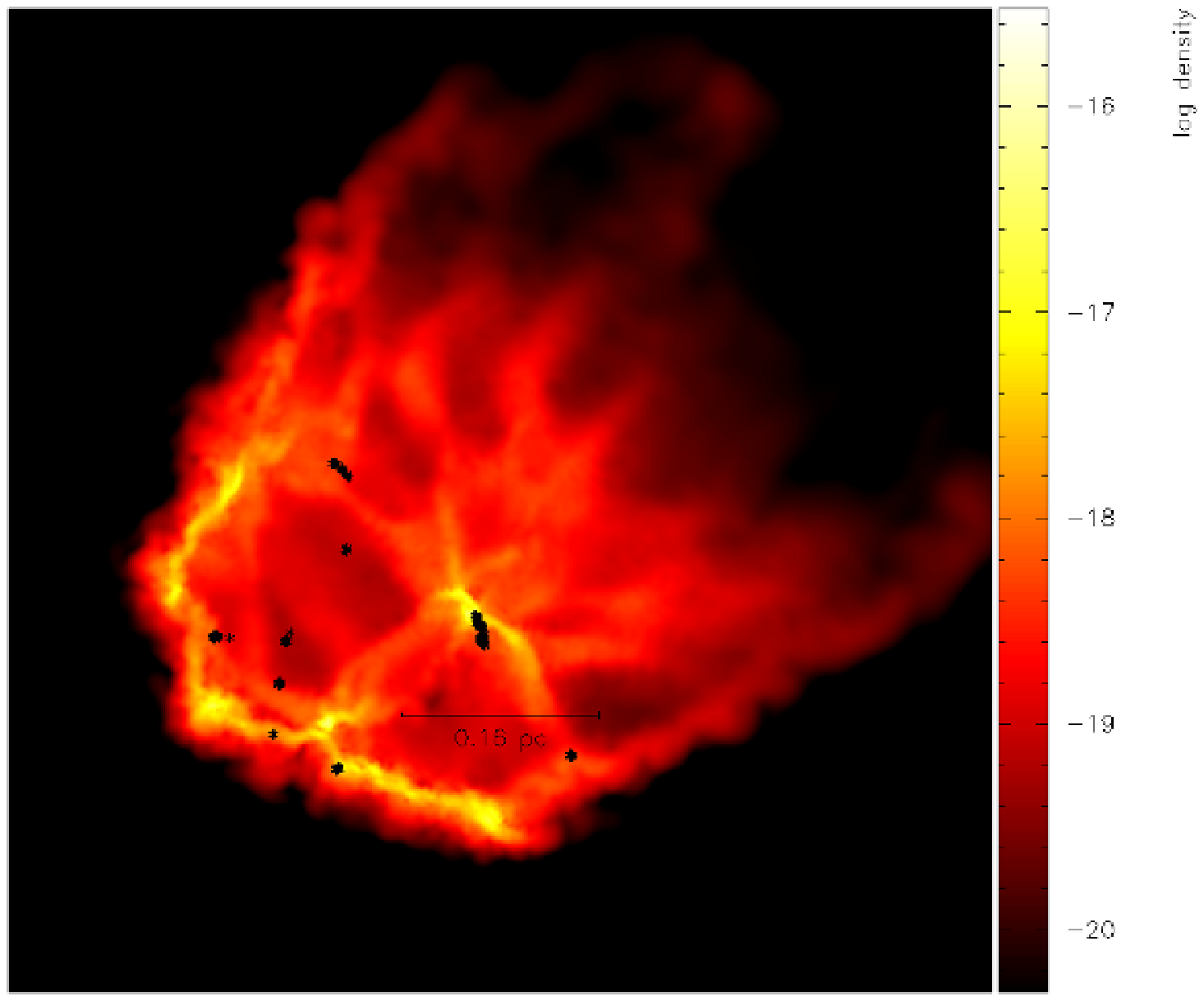}
 \includegraphics[angle=270, width=8.cm]{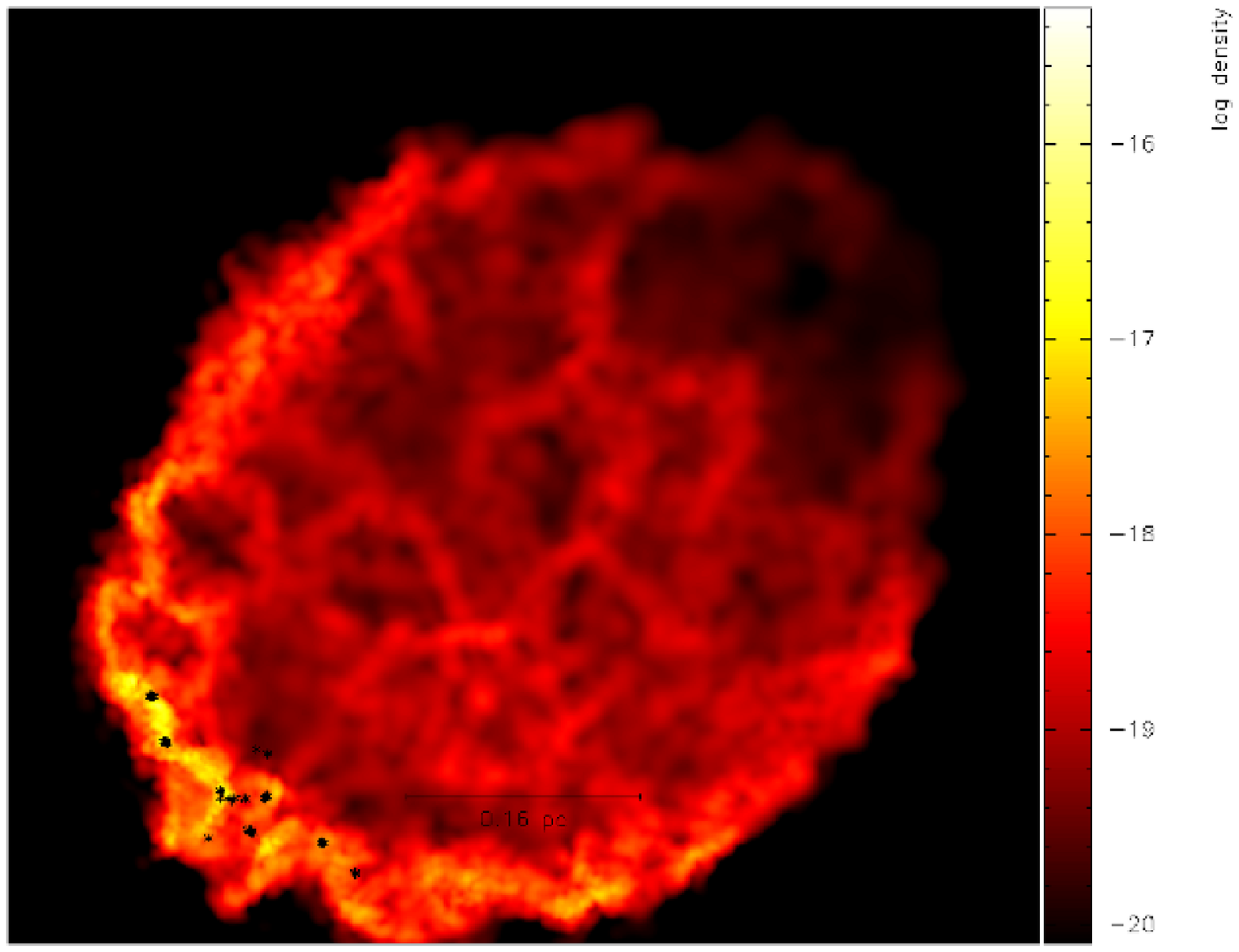}
 \caption{\emph{Left-panel}: A rendered column-density plot showing the mid-plane of the irradiated cloud in case 3 ($t\sim 0.75 t_{ff}$), after the formation of the first few protostellar objects. The protostars, represented by the sink-particles, have been marked with '$\ast$'. Observe that the formation of stars commences in the shocked layer on the cloud surface, though another potential site of star-formation appears in the central region of the irradiated cloud. \emph{Right-panel}: A rendered column density plot showing the mid-plane of the irradiated cloud in case 4. As in case 3, most of the star-formation in this case is also confined to the shocked layer on the irradiated surface of the cloud ($t\sim 0.1 Myr\equiv 0.33t_{ff}$). Shocked filaments within the cloud, thus far, do not appear to have acquired sufficient density to spawn stars. } \vfil} \label{landfig}
\end{figure*}

\section{Discussion}
Star-formation according to Krumholz \& McKee (2008), is unlikely to commence unless gas density is higher than a threshold. The proposition is attractive, for such a threshold will essentially reduce the problem of star-formation to simply an investigation into the processes likely to assist the assembly of a sufficiently dense volume of gas. However, it is common knowledge that the dynamical state of gas is also inextricably interwoven with its temperature, where for convenience, we include turbulence as just another thermal component. It is therefore difficult to define a density threshold for the formation of stars. 

There are two crucial issues before us : (i) the process, usually referred to as a trigger, that assembles pockets of dense gas, and (ii) the processes that maintain these pockets at the right temperature so they can spawn stars. Clearly, this latter question is beyond the remit of the present work, however, by maintaining isothermality within the cloud, at least the first half of the problem can be examined. Other crucial points that are also discussed while explaining the formation of stars are  of course, the timescale over which stars form which is also an indicator of their age, and the distribution of stellar masses at their birth, characterised by the initial mass function ({\small IMF}).

We now examine the results from our test simulations to find answers to these questions. While the turbulent velocity field originally introduced into the cloud produces dense filaments of gas, the effect of the incident flux of IR on the cloud depends on its strength as we have discussed so far. Firstly, all the 4 test simulations show that a suitable triggering mechanism enhances the rate at which gas is assembled in dense pockets. While some pockets of dense gas are isolated, yet others lie in larger filamentary structures. The strength of the trigger, however, determines the essential physical characteristics such as the mass of these pockets of dense gas and the timescale over which they are assembled within the cloud. 

\begin{figure}
 \vspace{12pt}
  \includegraphics[angle=270, width=9.cm]{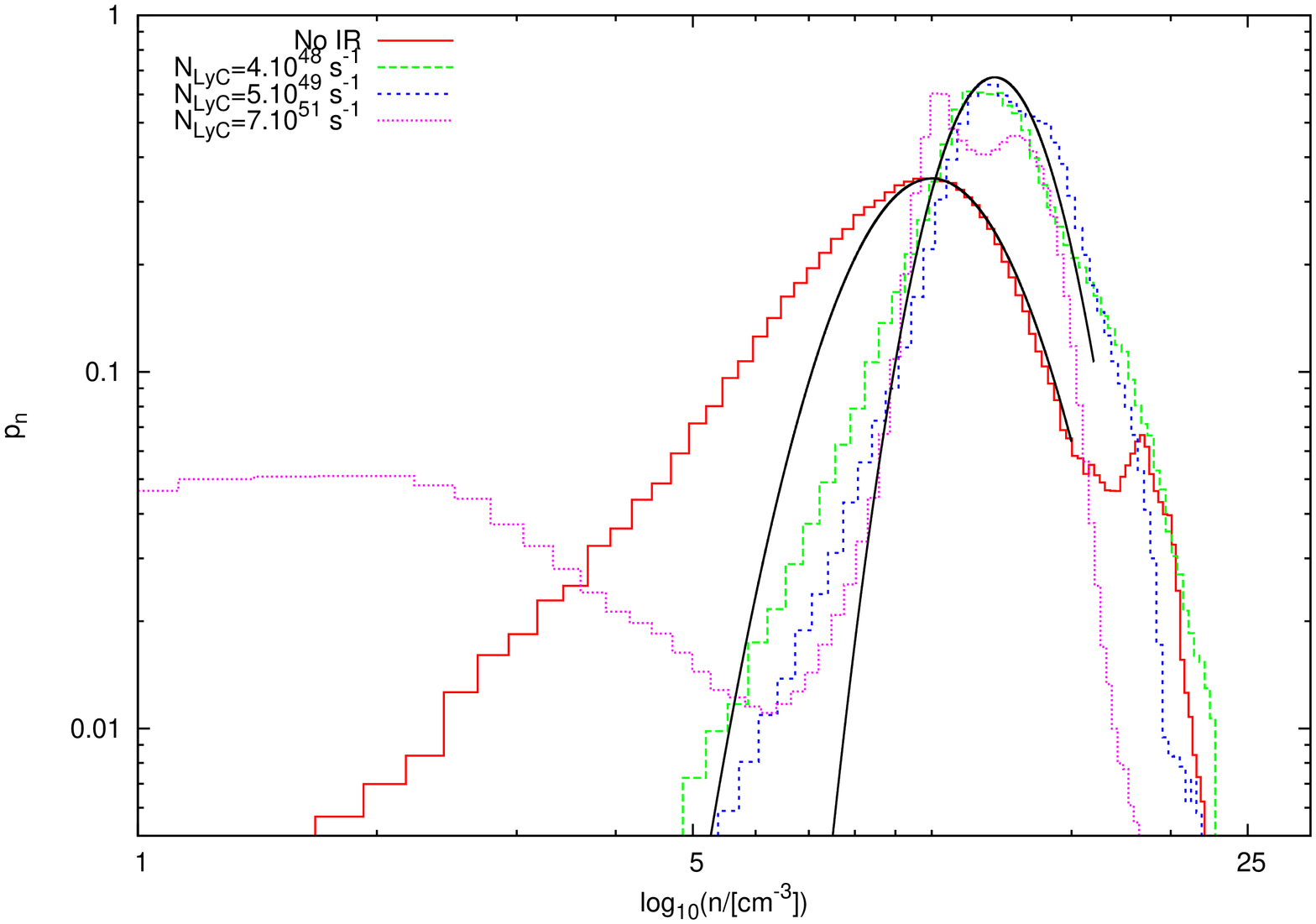}
  \includegraphics[angle=270, width=9.cm]{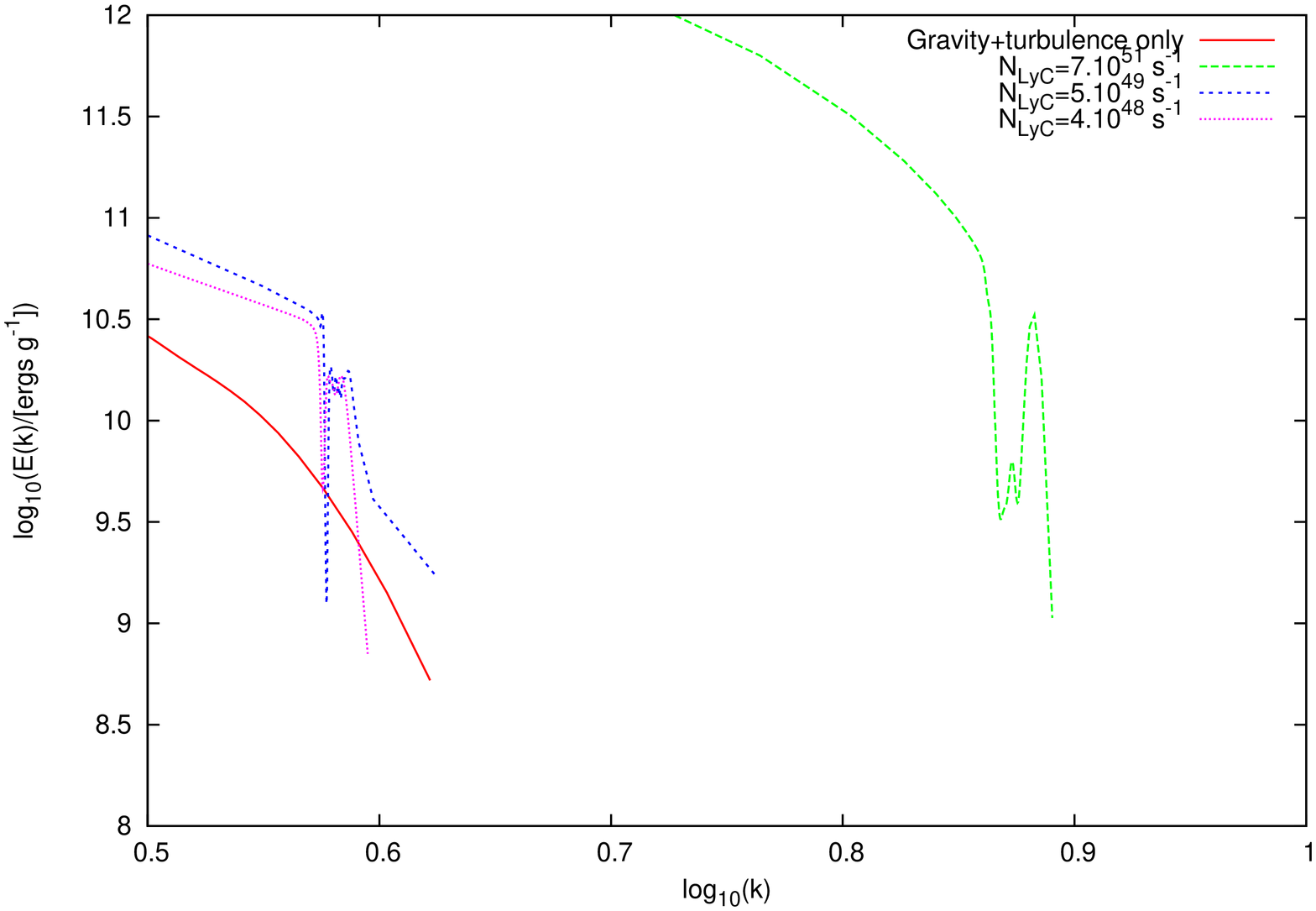}
 \caption{\emph{Upper-panel} : Figure showing the probability distribution function ({\small PDF}) for gas density within the cloud at an advanced stage of its evolution in each of the 4 test cases. A log-normal distribution fits all the {\small PDF}s reasonably well, though they also have a prominent power-law tail. We note that the width of the log-normal distribution is narrower for the irradiated clouds and the narrowest for the choice of the strongest flux. The best-fitting log-normal distributions have also been shown for comparative purposes. \emph{Lower-panel} : The power-spectrum for each of the 4 test cases has been plotted. The spectra show remarkable similarity to the Kolmogorov spectrum. We note that maximum power is cascaded down to smaller spatial scales in case 4.}
\end{figure}

The plots in Fig. 8 thus offer an instructive comparison. The retention factor, $f(\mathrm{M})$, plotted for the irradiated cloud in each test case shows that a progressively smaller mass of the cloud is retained with increasing strengths of the incident flux. The observed rate of photoablation in cases 3 and 4 is a few times $\sim 10^{-4}$ M$_{\odot}$ yr$^{-1}$, which is consistent with that calculated using Eqn. (4) above. Plotted in the lower-panel is the timescale on which gas within the cloud overhauls a fiducial density threshold, $\rho_{thresh}\sim 10^{-18}$ g cm$^{-3}$. Thus, with relatively weaker shocks resulting from damped turbulence in case 1, it takes considerably longer, more than 70 \% of the free-fall time, to overhaul a fiducial density threshold of $10^{-18}$ g cm$^{-3}$; see the plots in the lower-panel of Fig. 8.

 Exposure to a weak flux of IR as in test case 2 makes little difference to the rate at which gas overwhelms this threshold. Increasing the strength of the IR though strongly ablates the cloud and, more importantly, gas within the cloud also enters the dense phase on a relatively shorter timescale. This can be seen  easily  from the respective plots for cases 3 and 4 in the lower-panel of Fig. 8. The mass of gas within the irradiated cloud above the density threshold, M$(\rho_{thresh})$, rises most rapidly in case 4, where the IR-flux is the strongest. 

Our conclusions from the set of simulations discussed above are also corroborated by those reported recently by Walch \emph{et al.} (2012). These authors have drawn attention to the dependence on the fractal dimension of the evolution of an irradiated {\small MC}. They suggest two possible scenarios for a fractal cloud. Thus, a {\small MC} with a relatively large fractal dimension ($\gtrsim$2.6) was observed to produce pillar like features where as the cloud with a fractal dimension smaller than $\sim$2.2 became perforated as dense structure within the cloud was shocked by the incident radiation. To this end our results are consistent with this latter scenario; see the rendered plots in Figs. 6 and 9.

In another related work Dale \& Bonnell (2012), while reporting the findings from their simulations contradict these results, including ours. They found that the evolution of a turbulent {\small MC}, a few thousand M$_{\odot}$s massive, was largely unaffected even after exposing it to a  strong  flux of {\small IR}. Thus, irrespective of the flux of {\small IR}, the turbulent {\small MC} in their study ended up as a single central filament, although trunk-like features developed on the surface of this filament when exposed to {\small IR}. However, the resulting {\small MF}s were mutually comparable. Dale \& Bonnell (2012) therefore suggest that even a relatively strong flux of {\small IR} is unlikely to significantly alter the evolution and the star-formation history of that cloud.

 This radically different result is probably due to - (a) the difference in the modes in which turbulent energy  was initially injected, and (b) a relatively larger initial mass and density of their test cloud that could have impeded radiation-induced turbulence. The formation of a single filament in their simulations also suggests that the mode driven by the initial turbulence was dominant and grew rapidly before the energy injected by the {\small IR}-flux could have any impact. We now discuss the principle findings of our study.

\subsection{Probability density function ({\small PDF})}
The probability density function is a useful diagnostic to study the effects of physical conditions on the distribution of gas density within a {\small MC}. The {\small PDF}s for each of the 4 test cases have been plotted in the upper-panel of Fig. 11. In general, the {\small PDF}s appear to be a combination of power-law and lognormal distributions. It is evident that gas distribution within the irradiated cloud in cases 2 to 4 spans a wider rage of density as compared to the turbulent cloud in case 1 where it was not exposed to {\small IR}. The {\small PDF} in this latter case has distinct power-law tails at both, the low as well as high-density ends, albeit the power-law for  higher densities is much steeper. The distribution for intermediate densities though, is roughly lognormal as can be seen in the plot in the upper-panel of Fig. 11. The {\small PDF}s for the irradiated clouds in the remaining 3 cases are mutually similar, however, the power-law tail for each one of them is significantly steeper than that observed in case 1. The {\small PDF}s derived here and those reported by us for a shocked-cloud in an earlier work (Anathpindika \& Bhatt 2011), bear striking resemblance with those deduced by Schneider \emph{et. al.} (2012), for the Rosette {\small MC} irradiated by an adjacent cluster of stars, {\small NGC} 2244. These latter authors along with a few others (e.g., Hill \emph{et. al.} 2011), also suggest the possibility of the occurrence of more complex {\small PDF}s than those derived here. 

\subsection{Power-spectrum}
The spectra plotted for each of the four test cases and shown in the lower-panel of Fig. 11 resemble the Kolmogorov spectrum. The derived power-spectra  are also consistent with those for {\small MC}s in the Galactic ring (Roman-Duval \emph{et al.} 2011). Progressively increasing the strength of the turbulence both, drives a broader range of wave-numbers and produces a considerably steeper spectrum such as the one for case 4. Turbulent energy in these latter cases of radiation-driven turbulence cascades down to smaller length-scales which probably induces fragmentation on smaller spatial scales. The effect of the energy-cascade also manifests itself in the {\small PDF}s plotted in the upper-panel of Fig. 11 and discussed in the previous section. Enhanced fragmentation also shifts the protostellar mass-distribution towards lower masses as is indeed observed in the mass distributions derived here and discussed in \S 5.4.

\subsection{The star-formation rate: Is triggered star-formation rapid ?}
Having discussed one aspect of the problem, we now move to the next where the possible effects of shocks on the distribution of stellar masses are examined. The accretion history of sink particles in each simulation is plotted in Figure 12, where the combined mass of sink particles in each realisation has been shown. The spread along the abscissa reflects the distribution of ages of the sink particles in a particular test case. Importantly, the steep nature of the plots suggest that the formation of the first sink particle is followed by an episode of rapid sink-formation and a number of them form over the next few $10^{4}$ years.  Each simulation was terminated after the formation of 150 sinks though this means, the simulations were not coeval at the time of termination. We observe that sink-formation timescale in each test case is only a fraction of the age of the cloud itself, or indeed, its freefall time. 

The rate at which ordinary gas is converted into stars is quantified by the so called, star-formation rate ({\small SFR}). The {\small SFR} is also a measure of the efficiency with which an external trigger assembles potential star-forming sites. The star-formation rate ({\small SFR}), we remind, is actually the rate at which sinks form and has been calculated for each test case; see Table 2. It is the largest for case 1 where the turbulent cloud was left by itself to evolve under self-gravity. Irradiated clouds in cases  2, 3 and 4, on the other hand, show a mutually comparable, though relatively low {\small SFR}. However, these latter values of the {\small SFR} are consistent with those derived for typical star-forming clouds in the nearby universe (e.g. Lada \emph{et al.} 2012). The relatively small {\small SFR} in cases 2, 3 and 4 is not entirely surprising since injected turbulence causes sub-fragmentation of larger structures to produce smaller cores and therefore, less massive stars. This tendency is further reflected by the distribution of sink masses plotted in Fig. 13.

\begin{table}
 \centering
 \begin{minipage}{80mm}
  \caption{The observed rate of sink-formation for each test case ({\small SFR} = $\frac{\sum \mathrm{M}_{sinks}}{\mathrm{t}_{sinks}}$); $t_{sinks}$ marks the epoch when a simulation was terminated, in other words the age of the cloud in each case.}
  \begin{tabular}{lc}
  \hline
  \hline
Case &   {\small SFR}[$\times 10^{-4}$] \\
       &   [M$_{\odot}$ yr$^{-1}$] \\
\hline  
   1  & 5.61 \\
\hline
   2   &  1.07    \\
\hline
   3  & 1.42 \\ 
\hline
   4  & 1.94 \\
\hline
\end{tabular}
\end{minipage}
\end{table}

Authors have often cited similar values of the {\small SFR} to support the hypothesis of slow star-formation (e.g., Krumholz \& Tan 2007). Conversely, we have shown here that shocks triggered due to continuously driven turbulence rapidly assemble gas in the dense phase before eventually spawning stars. Stars thus form on a relatively short timescale, only of the order of a fraction of the free-fall time of the original cloud, though protostellar feedback may further reduce the {\small SFR} and so, the derived values here are likely tobe the upper limits. 

Recent studies of a number of star-formation sites such as those within the Rosette nebula (e.g., Wang \emph{et al.} 2008, Schneider \emph{et al.} 2012),  the {\small RCW}34 (Bik \emph{et al.} 2010), or the {\small SFO}38 (Choudhary \emph{et al.} 2010), have revealed several interesting properties, some of which can be reconciled with the numerical models discussed in this study. Yet in the well-known {\small MC}s such as the Orion and Monoceros, Wilson \emph{et al.} (2005) have reported a systematic increase in the age of stellar populations, indicative of sequential propagation of star-formation. Unfortunately, due to the absence of a feedback mechanism in our models we are ill-qualified to comment on this aspect of the star-formation process. At the least, we add that sequential propagation of star-formation is likely via feedback from the first generation of stars within a star-forming cloud.

\begin{figure}
\vspace {20pt}
 \includegraphics[angle=270, width=8.cm]{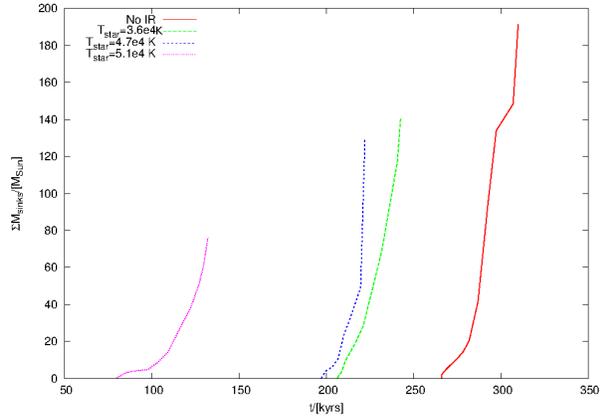}
 \caption{Plots showing the rate at which the total sink-mass in each of the 4 test cases increases. Observe that sink-formation in irradiated clouds occurs on a relatively short timescale; it is, however, significantly delayed in the cloud not exposed to the {\small IR}-flux where sinks appear only after a free-fall time ($t_{ff}\sim$ 0.27 Myr). }\label{landfig}
\end{figure}

\begin{figure*}
\vbox to 125mm{\vfil
 \includegraphics[angle=270, width=8.cm]{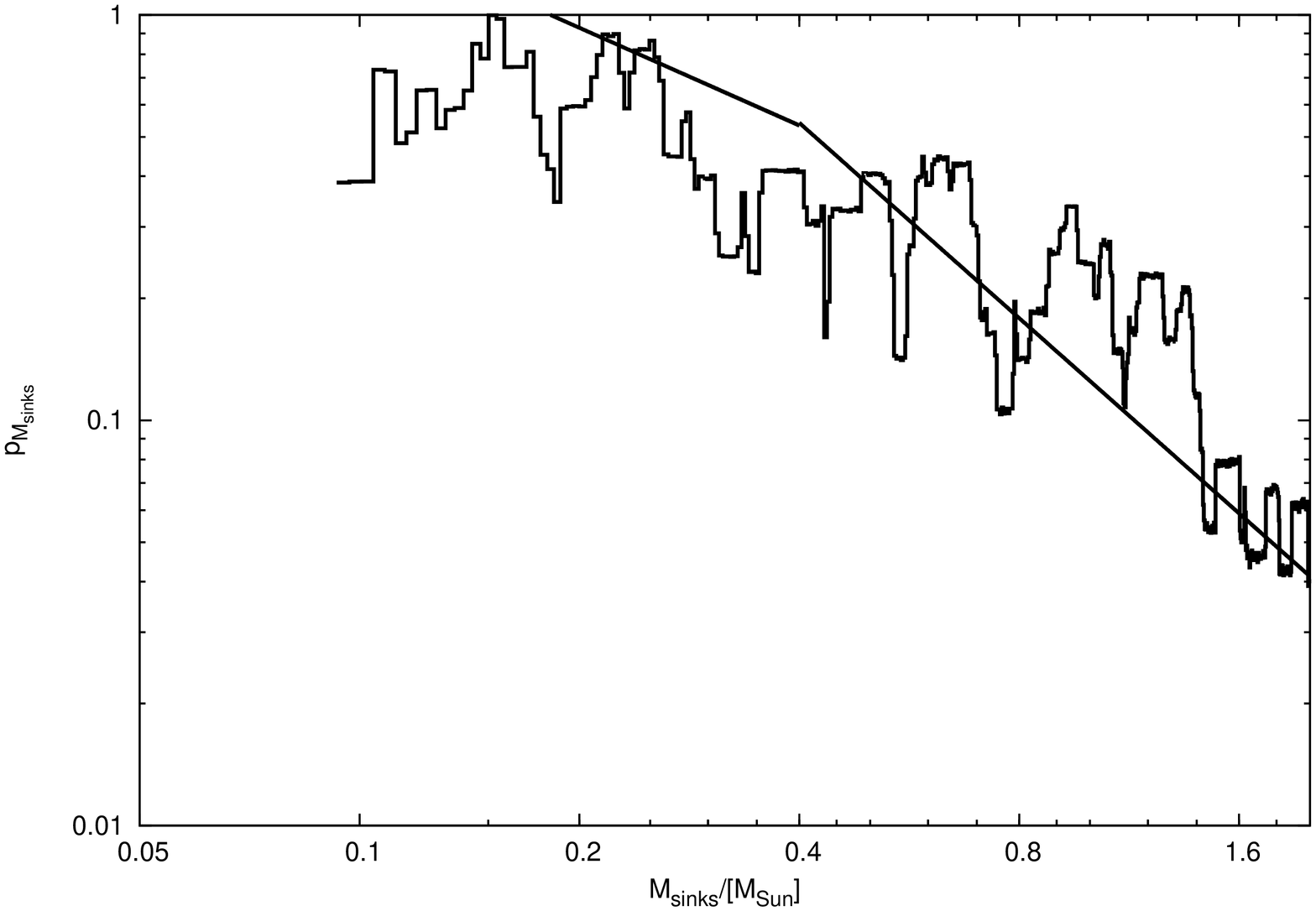}
 \includegraphics[angle=270, width=8.cm]{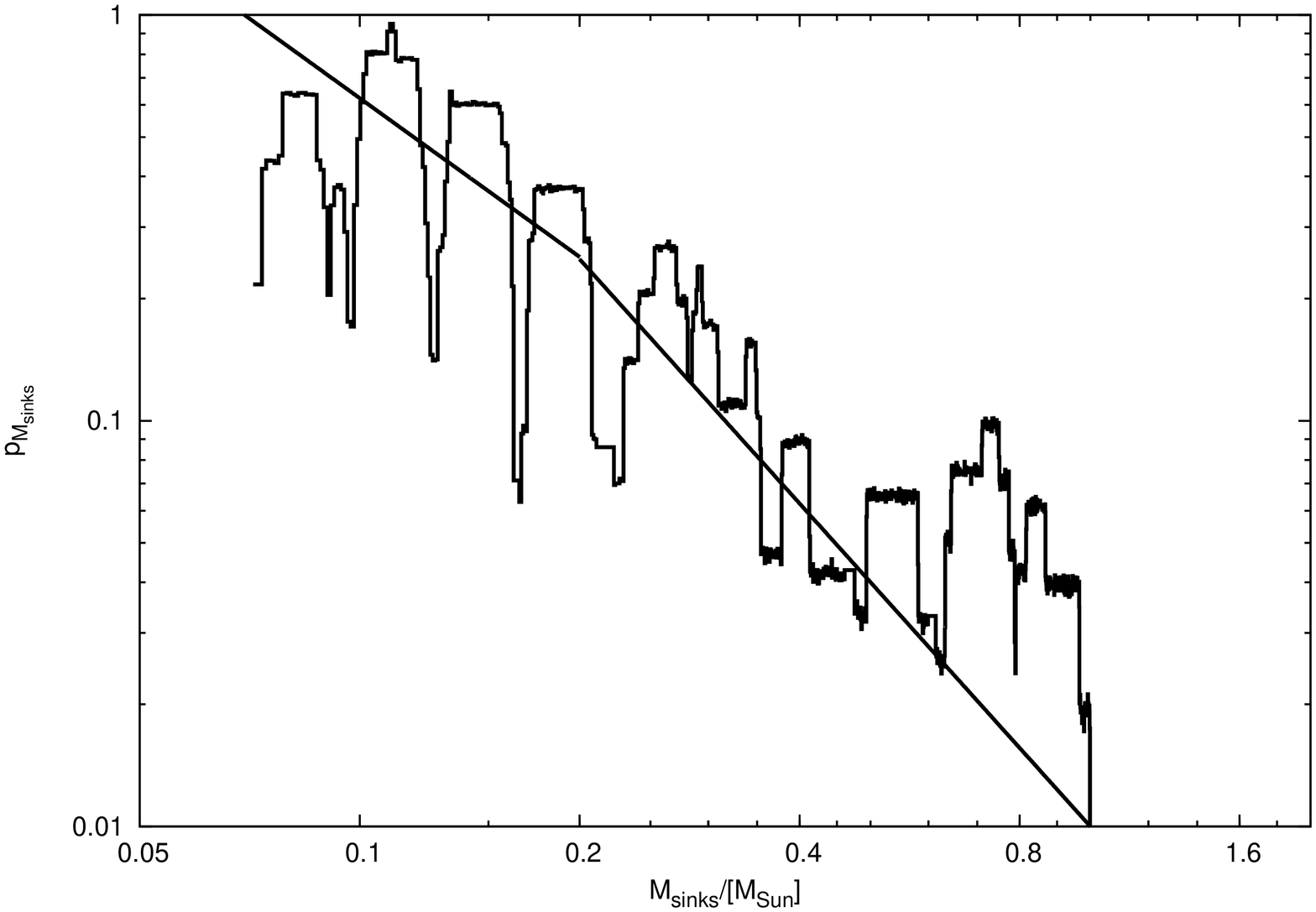}
 \includegraphics[angle=270, width=8.cm]{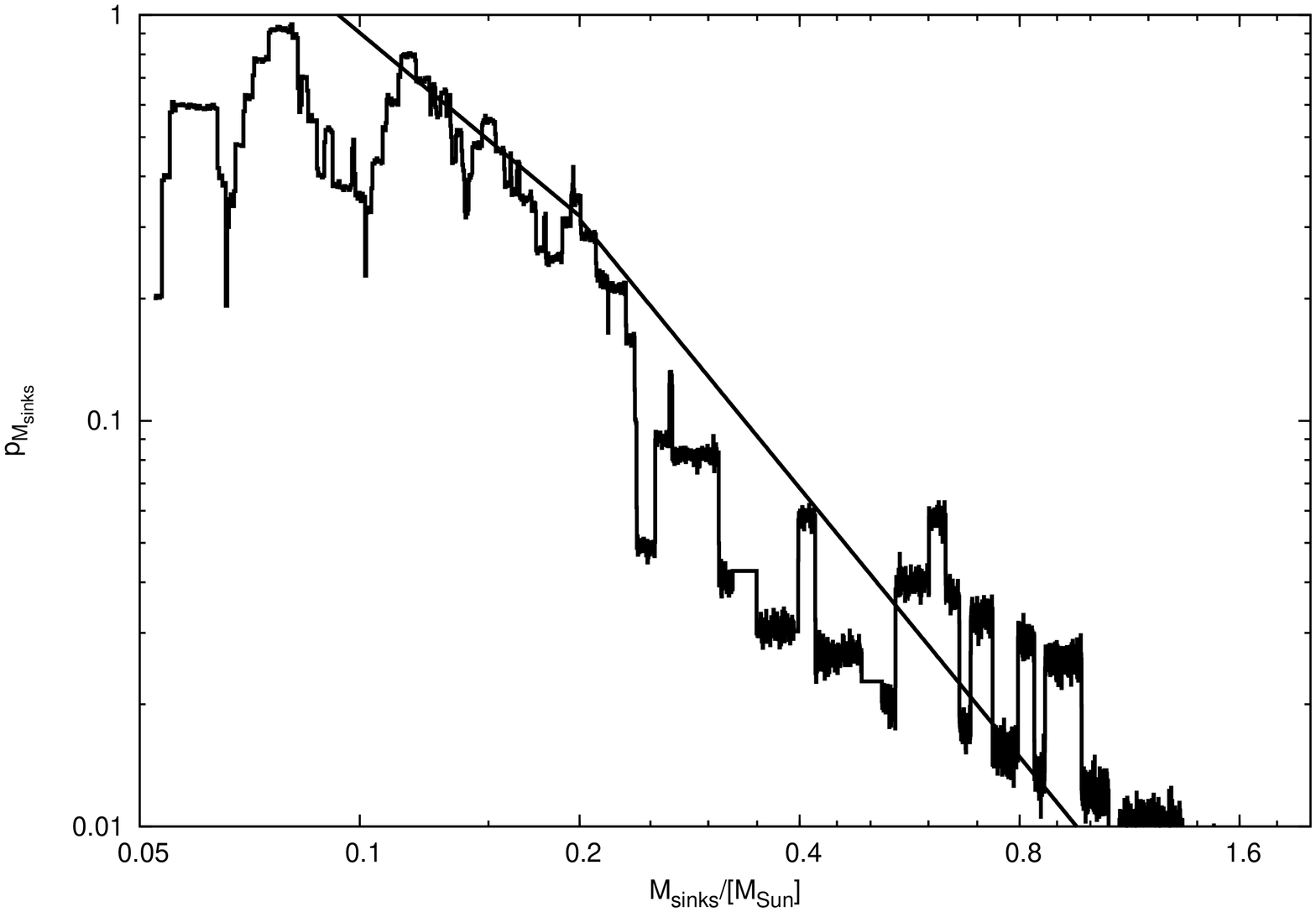}
 \includegraphics[angle=270, width=8.cm]{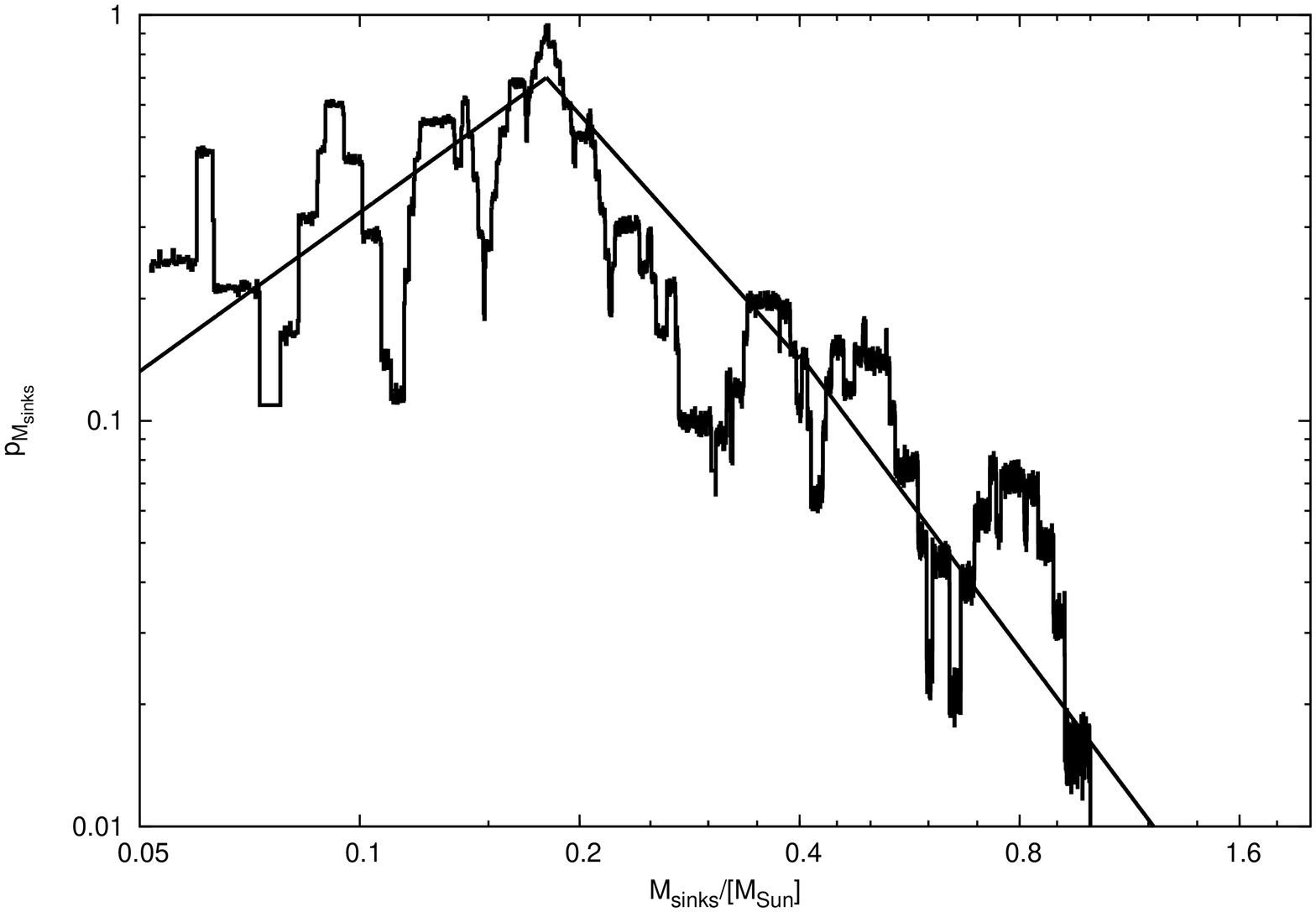}
 \caption{\emph{Top panel}: The sink mass function(MF) for cases 1 and 2 are shown in the left and right-hand plots. \emph{Lower panel}: Plots in the left and right-hand panels show the sink-mass functions for cases 3 and 4. Each MF was generated using masses for 150 sink particles. Note that MFs in the first 3 cases are incomplete below $\sim$0.05 M$_{\odot}$, where as in case 4, the MF turns-over towards masses smaller than $\sim$0.2 M$_{\odot}$.} \vfil} \label{landfig}
\end{figure*}

\subsection{Mass function ({\small MF})}
The mass function({\small MF}) generated using 150 protostellar objects (i.e., sinks) for each test case is shown in  Fig. 13. It will be useful to recall that the age of the fragmented cloud, and therefore the {\small MF}, at this stage is different in each test case, varying between $\sim$0.33 $t_{ff}$ in case 4, to $\sim$1.2 $t_{ff}$ in case 1. This can also be seen from the sink-accretion histories plotted in Fig. 12. Quite clearly, the {\small MF} in each test case has a power-law nature of the type
\begin{equation}
\frac{d\textrm{N}}{d\textrm{M}_{sink}} \propto \textrm{M}_{sink}^{-\alpha},
\end{equation}
for a non-zero $\alpha$. The parameters of the {\small MF}s derived for each case tested in this study are listed in Table 3. The {\small MF} for each case was derived by convolving 10$^{5}$ realisations of the respective data set within 1$\sigma$.
 Starting with a relatively shallow power-law in case 1 where turbulence is continuously damped, the {\small MF} steepens progressively as turbulent energy is continuously injected by increasingly stronger radiation-driven shocks in the remaining 3 cases. 

 The {\small MF} in case 4 where the flux of {\small IR} is the strongest, attains the canonical slope at the high-mass end. It develops a knee at $\sim$0.4 M$_{\odot}$, before turning over into lower masses at $\sim$0.2 M$_{\odot}$. The slopes, $(\alpha_{1},\alpha_{2})$, of this {\small MF} for masses $\lesssim$0.4 lie within the error-bars on the stellar initial mass function ({\small IMF}) due to Kroupa (2002). On the other hand, in cases 2 and 3 where the flux of {\small IR} is relatively weaker, the {\small MF}s develop a knee at $\sim$0.2 M$_{\odot}$. However, it shifts right-ward, i.e., to a slightly higher mass, $\sim$ 0.4 M$_{\odot}$, for case 1 where the derived {\small MF} itself is shallow and the cloud is not exposed to radiation. We note that the minimum mass, M$_{min}$, of an {\small SPH} particle defined in \S 3.3 above sets the threshold for the completeness of the MFs derived here. In the simulations discussed in the present work, M$_{min}$ $\sim$0.04 M$_{\odot}$, rendering the {\small MF}s incomplete below $\sim$0.05 M$_{\odot}$. The minimum mass, M$_{min}$, we remind, has been defined in \S 3.3 above as N$_{neibs}$\Big($\frac{M_{cld}}{N_{tot}}$\Big). It is therefore not possible for us to comment on the possibility of the formation of stars smaller than $\sim$0.05 M$_{\odot}$ via turbulent fragmentation.

\section{Conclusions}
In this article we presented a comparative study of the evolution of turbulent {\small MC}s. Turbulence, irrespective of whether it is damped or injected continuously, generates filamentary structure within {\small MC}s. In the first test case of this study, turbulence was allowed to damp without ever being replenished where as in the remaining three test cases, it was injected continuously via radiation-driven shocks. These latter three cases show that irradiating a large {\small MC} profoundly affects its evolution, and therefore, the star-formation history. The cloud with damped turbulence, however, evolves significantly differently compared to that in which turbulence is continuously injected via an external source. Not only is there a difference in the evolution, but the timescale on which the cloud evolves also varies and in fact, the cloud with stronger turbulence within it evolves much faster. Thus in the present work, the injected turbulence within the cloud is the strongest in case 4  and so, star-formation within the cloud in this case commences on the shortest timescale in comparison to the other 3 cases.

Global gravity soon dominates gas dynamics within the cloud evolving purely under self-gravity so that few modes other than the Jeans unstable mode are driven. Consequently the Jeans mode becomes the principal mode of instability so that gas within the {\small MC} is primarily channelled into this mode. Furthermore, this tendency is reflected by the {\small PDF} as well as the {\small MF} for the {\small MC} in case 1. Without much surprise, the {\small PDF} for this {\small MC} is largely predisposed towards somewhat higher masses, an obvious mis-fit to the canonical {\small IMF}. Driven turbulence, on the other hand, injects energy in to various unstable modes which aids formation of dense structure. In this study turbulence was injected within {\small MC}s by an uninterrupted source of {\small IR}. Simulations were performed by varying the strength of the {\small IR}-flux, spanning a range of radiation emitted by a typical {\small B}-type star to that emitted by young {\small OB}-associations. There is little doubt that progressively stronger radiation drives shocks of ever increasing strength. 

A common feature of the irradiated {\small MC}s observed in each of the three test cases, 2, 3 and 4 of this study, is the appearance of a shocked shell on the irradiated surface facing the source of IR and the shocked internal filamentary structure generated by the initial turbulence. The secondary shocks driven by the flux of {\small IR} create new sites of star-formation, particularly in the shocked shell and junctions of filaments within the {\small MC}. Evidently shocks distribute gas over a wider range of density as reflected by the corresponding {\small PDF}s. The range over which gas-density is distributed becomes wider with increasing strength of the radiation-driven shocks. Incidentally, these PDFs are consistent with those reported recently for star-forming clouds exposed to strong fluxes of {\small IR}.

Multiple stellar-systems appearing in isolated pockets of gas compete for gas as envisaged in the standard competitive-accretion scenario. The protostellar {\small MF}s derived in the present work, though fitted by a power-law, are not coeval. Interestingly, the {\small MF}s for the irradiated cloud in cases 2, 3 and 4 better agree with the canonical {\small IMF}, or even its more recent variation suggested by Kroupa. In fact, in case 4 where the radiation-induced shocks were the strongest, the {\small MF} even demonstrated a turnover at $\sim$0.2 M$_{\odot}$ which is consistent with the Jeans mass for the shocked gas maintained at its pre-shock temperature. The {\small MF} in case 1, though a power-law, is considerably shallow.Thus we have shown that cloud models with injected turbulence can possibly reconcile some key dynamical features of typical star-forming regions. This study, for instance, also shows that star-formation once triggered, proceeds rapidly and newer stars form within just over a few 10$^{4}$ years, which is significantly smaller than the free-fall time of a typical star-forming cloud. Also, the star-formation rate ({\small SFR}) derived for the irradiated cloud in cases 2, 3 and 4 is consistent with that reported for typical star-forming clouds. Thus, despite the dissipative nature of turbulence, it can profoundly affect the star-formation history of a {\small MC} and external sources such as a flux of {\small IR} can continuously replenish the turbulent energy within that cloud. Consequently, principle sources of turbulence such as  proto-/stellar feedback in nearby star-forming clouds, or nuclear star-formation activity in galaxies must be poignant for controlling the global star-formation history.

\begin{table}
 \centering
 \begin{minipage}{80mm}
  \caption{Physical parameters for the mass function (MF) derived in each test case. The indices $\alpha_{1}$ and $\alpha_{2}$ are the respective slopes on either side of the knee (i.e., M$_{sink}\ <$ M$_{knee}$ and  M$_{sink}\ >$ M$_{knee}$) of the MF. The MF only in case 4 turnsover for (M$_{sink}$/M$_{\odot})\lesssim 0.2$, and the corresponding slope is denoted by $\alpha_{3}$.}
  \begin{tabular}{lccccc}
  \hline
  \hline
Case & $\alpha_{1}$ & $\alpha_{2}$ & knee &  $\alpha_{3}$ & turn-over \\
      & & &   [M$_{\odot}$] & & [M$_{\odot}$] \\
\hline  
   1  & 0.9 & 1.6 & 0.4  &MF incomplete \\
      &&&& below $\sim$0.2 M$_{\odot}$\\
\hline
   2   & 1.3 & 2.0 & 0.2  & as above\\
\hline
   3  & 1.6 & 2.2 & 0.2 & as above\\ 
\hline
   4  & 2.0 & 2.4 & 0.4 & -1.3 & 0.2 \\
\hline
\end{tabular}
\end{minipage}
\end{table}

\section*{Acknowledgements}
  The authors are grateful to an anonymous referee for useful suggestions that improved the clarity of the original manuscript.
  Sumedh Anathpindika wishes to acknowledge post doctoral support provided by the Indian Institute of Astrophysics(IIA) at Bangalore. The simulations discussed in this work were performed on the HYDRA supercomputing cluster of the IIA. The author also thanks David Hubber and Thomas Bisbas for their support with {\small SEREN}. All rendered density plots in this paper were prepared using the publicly available visualisation package, SPLASH, prepared by Daniel Price.



%
%


\bsp

\label{lastpage}

\end{document}